\documentclass[12pt]{iopart}

\usepackage {hyperref}
\usepackage{graphicx}
\usepackage[dvipsnames]{xcolor}
\usepackage{iopams}

\begin{document}

\title[Nucleon-Nucleon correlations inside atomic nuclei]{Nucleon-nucleon correlations inside atomic nuclei: synergies, observations and theoretical models}

\author{Ranjeet Dalal$^1$, 
and I.J. Douglas MacGregor$^2$}
\address{$^1$ Guru Jambheshwar University of Science and technology, Hisar, India}
\address{$^2$ SUPA School of Physics and Astronomy, University of Glasgow, Glasgow G12 8QQ, United Kingdom}
\ead{ranjeet@gjust.org}

\
\vspace{10pt}
\begin{indented}
\item[]August 2022
\end{indented}

\begin{abstract}
While the main features of atomic nuclei are well described by nuclear mean-field models, there is a large and growing body of evidence which indicates an important additional role played by spatially-correlated nucleon-nucleon structures. The role of nucleonic structures was first suggested by Heidmann in 1950 to explain pick-up reactions of energetic nucleons. Since then a steady flux of new experimental evidence has confirmed the presence of similar structures inside atomic nuclei, dominated by correlations between pairs of nucleons. The role of these internal nucleon-nucleon correlations has been established using various energetic probes like photons, pions, leptons and hadrons. These correlated structures are essential for understanding the interaction of particles with nuclei and their presence provides an explanation of many specific nuclear phenomena including backscattered protons, copious deuteron production, sub-threshold particle production, neutrino interactions with nuclei and the EMC effect. On the theoretical side, these measurements have stimulated a large number of phenomenological models specifically devised to address these enigmatic observations. While reviews exist for specific interactions, there is currently no published commentary which systematically encompasses the wide range of experimental signatures and theoretical frameworks developed thus far. The present review draws together the synergies between a wide range of different experimental and theoretical studies, summarises progress in this area and highlights outstanding issues for further study.
\end{abstract}

\maketitle

\section{Introduction}\label{intro}

The internal structure of nuclei has traditionally been investigated using a wide range of energetic probes, leading to the ejection of one or more nucleons, or to meson production. These probes include photons \cite{Mac}, pions \cite{Weyer}, electrons \cite{Hen}, protons \cite{Pia}, neutrons \cite{Riley} and neutrinos \cite{Katori}. The interaction of probes with nuclei have provided vital information about the state of the nucleus and individual nucleons just before the interaction.\\

The interaction of energetic photons with nuclei was initially expected to result in the ejection of individual nucleons \cite{Courant} from specific shell model orbits. However, the observed nucleon angular distributions, cross-sections and the ratio of ejected protons and neutrons measured in early experimental work, led Levinger \cite{Lev1} to conclude that in many cases more than one nucleon was involved in the interaction. To explain these unexpected observations, he formulated the simple quasi-deuteron model (QDM) in which neutrons and protons often occur in deuteron-like structures inside nuclei. The resulting model, which is at odds with the nuclear shell model, has been quite successful in explaining the various subtleties of photon interactions with nuclei for $E_\gamma$ above 40 MeV \cite{Mc, Odian, Stein, Wat, Lep, Smith}. In this quasi-deuteron process, the photon-energy absorption predominantly takes place on a strongly interacting neutron-proton pair while the remaining nucleons act as spectators, making the $(\gamma,np)$ reaction an attractive tool to investigate 2N correlations in nuclei. \\

The dominance of quasi-deuterons, or spatially correlated neutron-proton structures inside nuclei, was also confirmed by measurements of charged-pion absorption in nuclei \cite{Weyer}, where absorption on two-nucleon pairs has undergone extensive testing since the 1950s. \\

Note that the nature of the correlation, or interaction, between the interacting nucleons depends on their spatial separation. At long range it is often modelled by single-pion exchange mechanisms, while at shorter distances it may involve multiple meson exchange or direct quark or gluon exchanges, if the nucleons are close enough to overlap. In general, higher energy probes are required to study shorter spatial separations and the commonly used term Short Range Correlations (SRC) is reserved for correlations studied with such high energy probes.\\

Recently, the inclusive and exclusive quasi-elastic reactions $(p,p{}^{\prime}NN)$ and $(e,e{}^{\prime}NN)$ have been investigated at very high momentum transfer at BNL \cite{Pia} and Jefferson Lab \cite{Hen}, respectively, where the angular correlations between ejected nucleons have been studied to reveal the state of nucleons just prior to the interaction of energetic probe particles. For ejected nucleons with momenta above 250 MeV/c, it has been observed \cite{Pia} that around 92$\%$ of the time protons ejected from ${}^{12}$C targets are accompanied by correlated neutrons. The most likely explanation is that the emitted high-momentum nucleons originate from short-range spatially correlated nucleon pairs. Such SRCs involving neutrons and protons are similar in construct to the quasi-deuteron concept first formulated to describe lower energy interactions. \\

The presence of a significant fraction of nucleons in strongly correlated 2N clusters, mainly quasi-deuterons or neutron-proton SRCs, leads to many other subtle observations including enhanced deuteron ejection by energetic probes. Pion absorption reactions \cite{Weyer} have led to copious deuteron ejection (up to 25$\%$ compared to the yield of ejected protons or neutrons) from nuclei. Similar observations were reported by Baba \textit{et al.} in 1984 for photons of multi-hundred MeV energies \cite{Baba}. A significant fraction of energetic photon and pion absorption events result in either the predominant and simultaneous ejection of kinematically correlated proton-neutron pairs, or the emission of deuterons. \\

The presence of nucleon-nucleon correlations inside nuclei was first used to explain the properties of pickup reactions by energetic neutrons \cite{Heid}. A notable example is copious deuteron ejection observed at forward angles by multi-hundred MeV hadrons, reported in early experiments by Azhgirei \textit{et al.} \cite{Az}. This was understood in terms of underlying quasi-deuteron degrees of freedom inside nuclei by Sutter \textit{et al.} \cite{Sutter}. These observations have been confirmed by recent experiments where the coincidence measurement of elastically scattered protons and deuterons are used to investigate the spin-isospin character of SRCs at Osaka \cite{Tara}. The unexpectedly high deuteron production, observed in numerous experiments using collisions of multi-GeV relativistic protons (for example, the early work of Cocconi \textit{et al.} \cite{Cocconi}), may be seen as extension of Azhgirei and Sutter’s quasi-deuteron propositions. \\

There have been many other independent manifestations of few-nucleon quasi-bodies inside nuclei, for example neutrino interactions with nuclei \cite{Katori,Ivanov}, including the significant fraction of 2-proton emission events \cite{Acciarri}, the observation of backscattered protons with unexpectedly high energy for multi-hundred MeV proton beam \cite{Haneishi,Miake}, the sub-threshold production of particles \cite{Knoll, Muller} and the EMC effect \cite{Weinstein}.\\

On the theoretical side, there have been intense activities to explain unexpected observations, most notably the development of the independent pair model or IPAM \cite{Gomes} and through the high-momentum components of nucleons in an independent particle framework \cite{Hen}. The other notable models considering SRCs inside nuclei are the unitary correlation operator method (UCOM) \cite{Roth} and recently proposed contact formalism \cite{Weiss}. The present work is an effort to encompass a wide range of experimental investigations and theoretical works illuminating the correlated structures observed within nuclei. A holistic review may stimulate further investigations of these phenomena. \\

\section{Experimental Indicators}\label{expt}

The first evidence for nucleon-nucleon correlated structures inside nuclei appeared shortly after the availability of energetic beams of photons, neutrons and pions more than 70 years ago. The range of evidence supporting spatially-correlated nucleonic structures has grown considerably since then. In this section, some of the key experimental indicators of strong nucleon-nucleon correlations are discussed. Our intention is not to present an exhaustive review of all the available evidence, but to underscore the ubiquitousness of correlations observed through a wide and diverse set of experiments. 
The most important observations advocating correlated degrees of freedom are: 

\begin{enumerate}
\item	Interaction of energetic photons with nuclei through the quasi-deuteron mechanism;
\item	Inclusive and exclusive reactions $(e/p, e{}^{\prime}/p{}^\prime NN)$ using energetic electrons/
protons; 
\item Pion absorption reactions where absorption of pions takes place mainly through 2N-absorption mechanisms;
\item	Direct quasi-elastic knockout of deuterons at forward angles using nucleons of a few hundred MeV energy; 
\item Backscattering of hadrons of several hundred MeV energy from nuclei and sub-threshold production of energetic particles; 
\item	Interaction of neutrinos with nuclei;
\item	The EMC effect.
\end{enumerate}

\begin{figure}[htbp]
  \centering
  \includegraphics[scale=0.7]{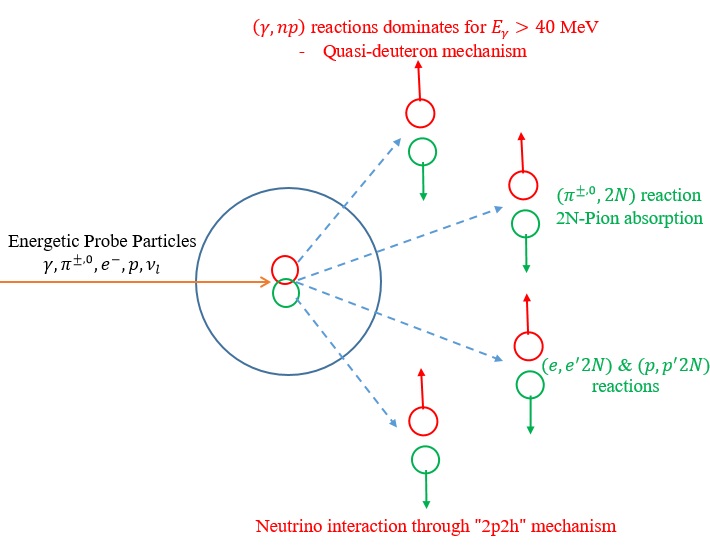}
  \caption{Direct demonstration of strongly-correlated 2N clusters (mainly in form of n-p pairs) by different incident particles.}
  \label{fig-1}
\end{figure}

\begin{figure}[htbp]
  \centering
  \includegraphics[scale=0.45]{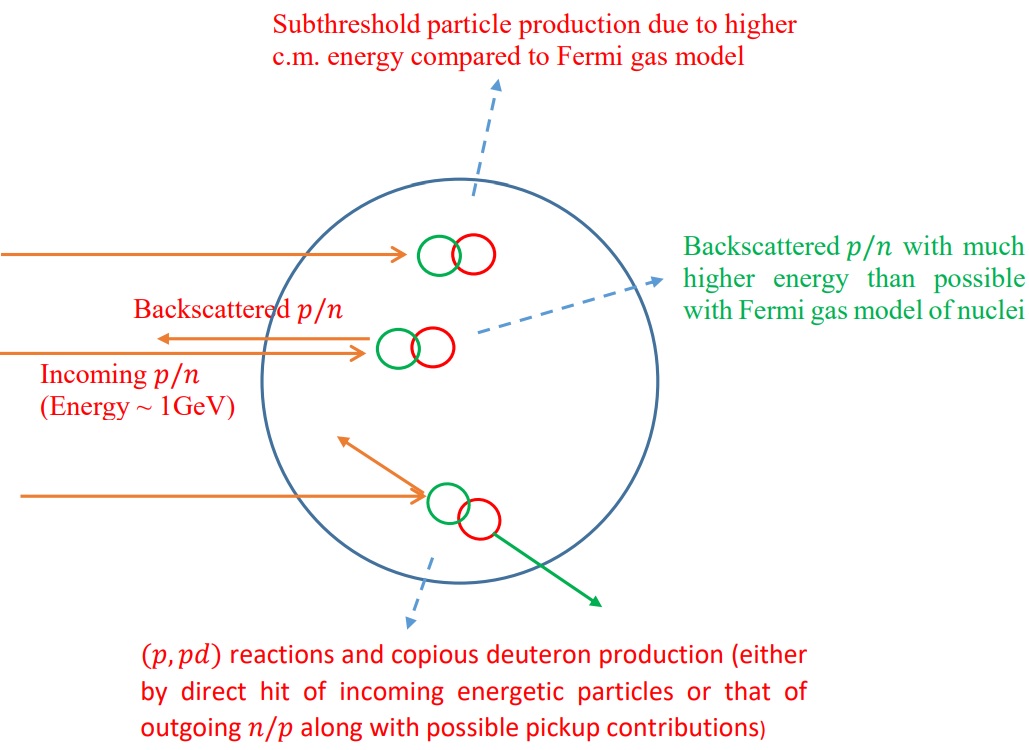}
  \caption{Other direct and indirect manifestations of strongly-correlated 2N systems inside nuclei.}
  \label{fig-2}
\end{figure}

A schematic illustration of these interactions is shown in figures \ref{fig-1} and \ref{fig-2}.  

\subsection{Interaction of energetic photons $(E_\gamma > $ 40 MeV) with nuclei}\label{src1}

In photoemission reactions, the outgoing nucleons were expected to follow either the direct photoelectric process developed by Courant \cite{Courant} in which photon energy is absorbed by individual proton leading to its ejection from nucleus, or the statistical theory of Blatt and Weisskopf \cite{Blatt} in which nucleons are assumed to “boil off” from the excited compound system. However, it rapidly became apparent that the photo-neutron and photo-proton production cross sections, and their ratio, could not be accounted for by either of these reaction models. It became possible to investigate the photo-disintegration of nuclei with much higher photon energies with the advent of bremsstrahlung photons from synchrotrons. It was expected that only nucleonic degrees of freedom would be important for $E_\gamma \sim $ 100 MeV, and  the $(\gamma , p)$ reaction would be much more probable due to its strong electric dipole component. In comparison the $(\gamma , n)$ reaction has a much weaker magnetic dipole contribution. Furthermore, no significant angular correlations were expected between ejected nucleons for reactions in which more than one nucleon were “boiled off”. However, energetic gamma-ray interactions resulted in comparable $\sigma_{(\gamma,n)}$  and $\sigma_{(\gamma,p)}$ cross-section values which were much higher than the single particle cross sections expected from Courant’s work.\\

This led Levinger \cite{Lev1} to propose a phenomenological quasi-deuteron model (QDM) in which protons and neutrons are assumed to form deuteron-like spatial sub-structures inside bound nuclei. The model was constructed on similar lines to the ideas employed by Heidmann \cite{Heid} to explain pickup reactions with energetic neutrons (90 MeV). The wavefunction for the ground-state of target nucleus ${}^A_Z X$ is assumed as product of wavefunction of a quasideuteron and a spectator part of nucleus with A-2 nucleons which do not participate in photo-disintegration process,\\
\begin{eqnarray}
 \psi (1,2,3,...A) = \exp(i k^\prime\cdot r^\prime) \psi_k(r) \phi (3, ....A). \label{eq1} 
\end{eqnarray}
\\
Here, $\exp(i k^\prime\cdot r^\prime)$ represents the centre-of-mass (CM) movement of quasi-deuterons with wavefunctions $\psi_k(r)$. The remaining portion of A-2 spectator nucleons constitutes a potential well containing the quasi-deuteron and may scatter/absorb the outgoing nucleons, an effect which is generally neglected for light nuclei. For any nucleus ${}^A_Z X$, the QDM cross section $\sigma_{X(\gamma,pn)} $ would be proportional to product of the photo-disintegration cross section for the free deuteron $\sigma_{d(\gamma,pn)} $ and the possible number of quasi-deuterons in the target nucleus (NZ/A), i.e.\\
\begin{eqnarray}
\sigma_{X(\gamma,pn)} = L \ \frac{NZ}{A} \ \sigma_{d(\gamma,pn)}. \label{eq2} 
\end{eqnarray}

Where L is Levinger’s proportionality constant. \\

The QDM predicted strong angular correlations between ejected neutrons and protons (back-to-back in the interaction CM frame). This was investigated in many experiments for a range of light and intermediate mass nuclei \cite{Odian,Stein,Wat} within a few years of the proposed model. These experiments established the validity of the quasi-deuteron approach for $E_\gamma >$ 40 MeV. In one early investigation \cite{Wat}, coincidence measurements of outgoing protons and neutrons were carried out for many target nuclei at the MIT synchrotron with a 340 MeV bremsstrahlung beam. The measurements used a fixed proton telescope at 76${}^\circ$ (corresponding to 90${}^\circ$ CM) for 260 MeV photons, while the angle of the neutron detector was varied. The proton detector was fitted with a suitable absorber such that only protons with energies greater than 120 MeV were detected. The angular distribution of neutrons for neutron-proton coincidence measurements on the deuteron, Li, C, O, Al and Cu are shown in figure 3. The neutron distributions for Li, C, O, Al and Cu are similar to that from the deuterium target, validating the basic assumption of the QDM. Further, broader neutron-proton coincidence peaks, compared to the ${}^2$H target, are obtained for these nuclei reflecting the CM momentum of quasi-deuterons inside nuclei.   Since, the initial energy and photon momentum is shared with close spatially correlated neutron-proton/quasi-deuteron pairs they are simultaneously ejected from nuclei resulting in similar cross-section values for $(\gamma,n)$ and $(\gamma,p)$ reactions. The ejected neutrons and protons are observed to have high relative momentum, and low CM momentum which may be fitted with a Gaussian distribution \cite{Wat}. \\

\begin{figure}[htbp]
  \centering
  \includegraphics[scale=0.8]{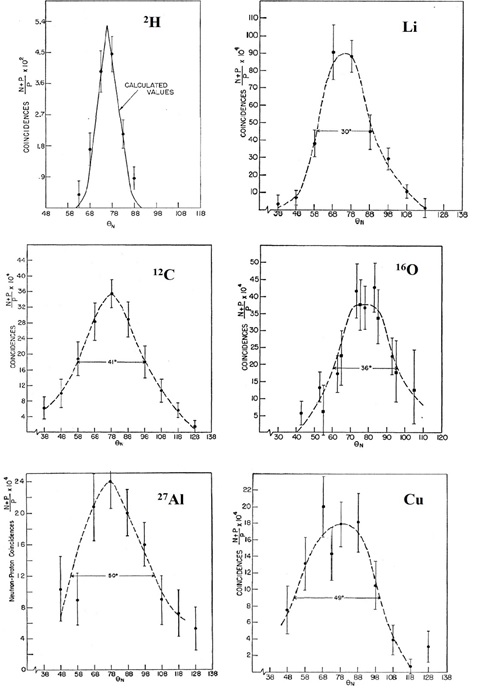}
  \caption{Neutron-proton coincidence measurements for ${}^2$H, Li, ${}^{12}$C, ${}^{16}$O, ${}^{27}$Al and Cu targets using 340 MeV Bremsstrahlung beam. The ratio of coincidences to the total number of single protons is plotted as a function of the angle of the neutron detector. The finite opening angles for neutron and proton detectors $(\sim$ 10${}^\circ$ each), beam profile and target thickness are major contributors to the error bars \cite{Wat}. The dashed line is drawn to connect the experimental points.}
  \label{fig-3}
\end{figure}

Soon, additional modifications were proposed to the QDM. The outgoing nucleons can have secondary interactions with other nucleons in the residual system \cite{Stein} leading to a reduction in correlated neutron-proton numbers. The probability of escape P(2R/$\lambda$) for neutron-proton pairs depends on the nuclear radius and mean free path of nucleons in the residual system \cite{Stein}. The secondary interactions, or Final State Interactions (FSI), lead to significant multi-nucleon ejections in the case of mid-mass and heavy nuclei \cite{Lep}. The investigation of the ${}^4$He$(\gamma,p) {}^3$H reaction \cite{Schmieden} using tagged photons of energy 80-160 MeV established that a significant fraction of $(\gamma,p)$ events originated from quasi-deuteron photo-absorption followed by neutron reabsorption. Incidentally, this also explained the observation of much higher cross-sections for $(\gamma,n)$ processes compared to those expected from traditional quasi-free knockout. \\

The value L = 6.4 is considered in the present work, though L values from 4 to 10 have been reported in the literature \cite{Smith}. The strong Coulomb barrier in the case of heavy nuclei would lead to an additional reduction in number of ejected protons. A damping factor $ exp{(-D/{E_\gamma})}$  with $D = 60$ MeV was introduced by Levinger \cite{Lev2}, to account for the lower $(\gamma,np)$ experimental cross sections, particularly at lower photon energies, compared to earlier predictions. This factor was later replaced by Chadwick \cite{Chadwick} using arguments in support of a Pauli-blocking factor. Irrespective of these various attempts to refine the proportionality constant, the persistence of the scaling with the deuterium cross section is by far the most striking and important feature of the QDM.\\

\begin{figure}[htbp]
  \centering
  \includegraphics[scale=0.45]{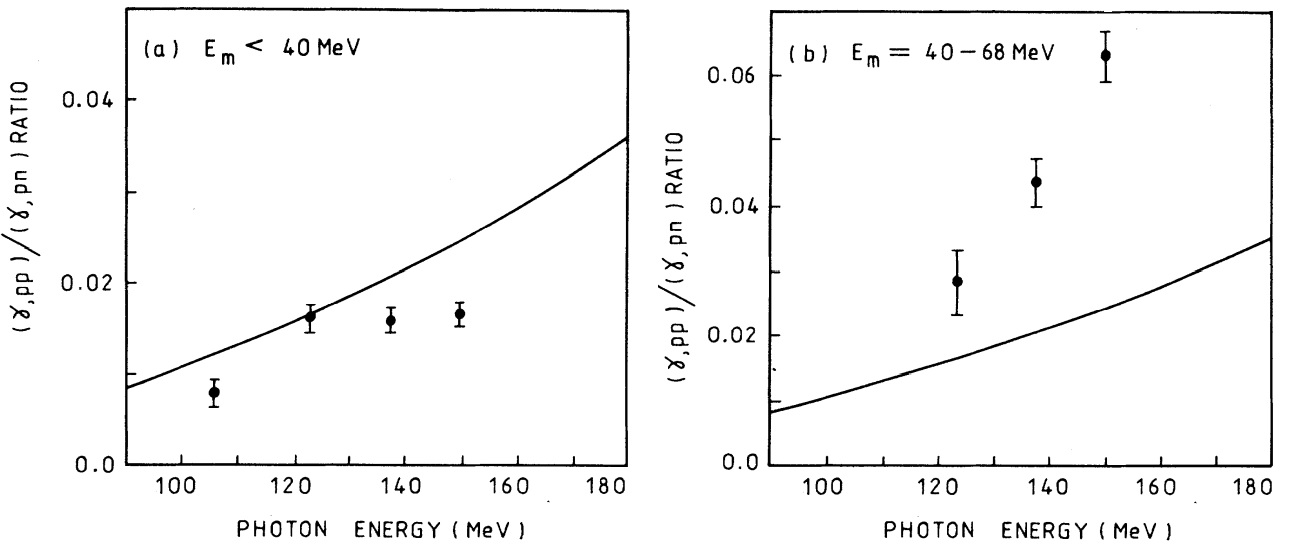}
  \caption{The cross-section ratio of $(\gamma,pp)/(\gamma,pn)$ reactions for ${}^{12}$C target nuclei for two different missing energy ranges; up to 40 MeV missing energy (left plot) and for missing energy of 40 to 68 MeV (right plot) \cite{Mc}. The theoretical calculation by Ryckebusch \textit{et al.} \cite{Ryckebusch} is depicted by the solid line}
  \label{fig-4}
\end{figure}

A systematic study of photon interactions with the ${}^{12}$C nucleus was later carried out by McGeorge \textit{et al.} \cite{Mc} using the tagged photon technique, for $E_\gamma$ = 80-157 MeV. In this energy region, the $\gamma$-interaction resulted in a predominately back-to-back emission of protons and neutrons in the CM frame and was well described by a quasi-deuteron mechanism. The observed $(\gamma,pp)/(\gamma,np)$ ratio was about 2$\%$ at low missing energy (up to 40 MeV) – a value which increased with the missing energy of reaction products suggesting that many of the $(\gamma,pp)$ events at higher missing energies may originate from initial $(\gamma,np)$ events followed by FSI of outgoing neutrons. A plot of $(\gamma,pp)/(\gamma,np)$ cross-section-ratio for  ${}^{12}$C is shown in figure \ref{fig-4} \cite{Mc}. Moreover, the momentum distribution of proton-neutron pairs inside nuclei was extracted and observed to peak, as expected, around 120 MeV$/$c. An extended summary of the programme of photonuclear experiments carried out at Mainz can be found in a recent reports by MacGregor  \cite{Mac, Mac2}. \\

The suppression of the quasi-deuteron mechanism for $E_\gamma <$ 40 MeV demands some scrutiny. As observed by McGeorge \textit{et al.} \cite{Mc}, the $(\gamma,pn)$ events in ${}^{12}$C are accompanied with at least 20 MeV missing energy $(E_m)$. That is due to the binding energy of two outgoing nucleons and excitation energy of residual system. For $E_\gamma$ below 40 MeV, the outgoing neutrons and protons would be left with very low kinetic energies ($\sim$10 MeV or less) which would be difficult to observe. At such low energies it is also likely that outgoing nucleons, particularly protons, would have small transmission factors. Moreover, the cross sections of single-charge exchange (SCX) $(n,p)$ or $(p,n)$ for different nuclei vary strongly with the energy of outgoing nucleons and generally peak for kinetic energies of protons/neutrons below 25 MeV. For example, the excitation function of ${}^{65}$Cu$(p,n)$${}^{65}$Zn peaks at about 12 MeV with $\sigma \sim$750 mb which quickly reduces to $\sigma \sim$ 25 mb at 25 MeV \cite{Kumar}. These factors would lead to a significant reduction in the QDM cross-section measured by coincidences between outgoing neutrons and protons at low photon energies. On the other hand, indirect $(\gamma,n)$ measurements \cite{Vey} indicate that the $(\gamma,np)$ reaction contribution starts just above the energy threshold for the reaction. In a study by Veyssiere \textit{et al.}  \cite{Vey}, $\beta$-activities of residual nuclei were used to estimate the cross-section for $(\gamma,pn)$ reactions inside the GDR region.\\

\begin{figure}[htbp]
  \centering
  \includegraphics[scale=0.45]{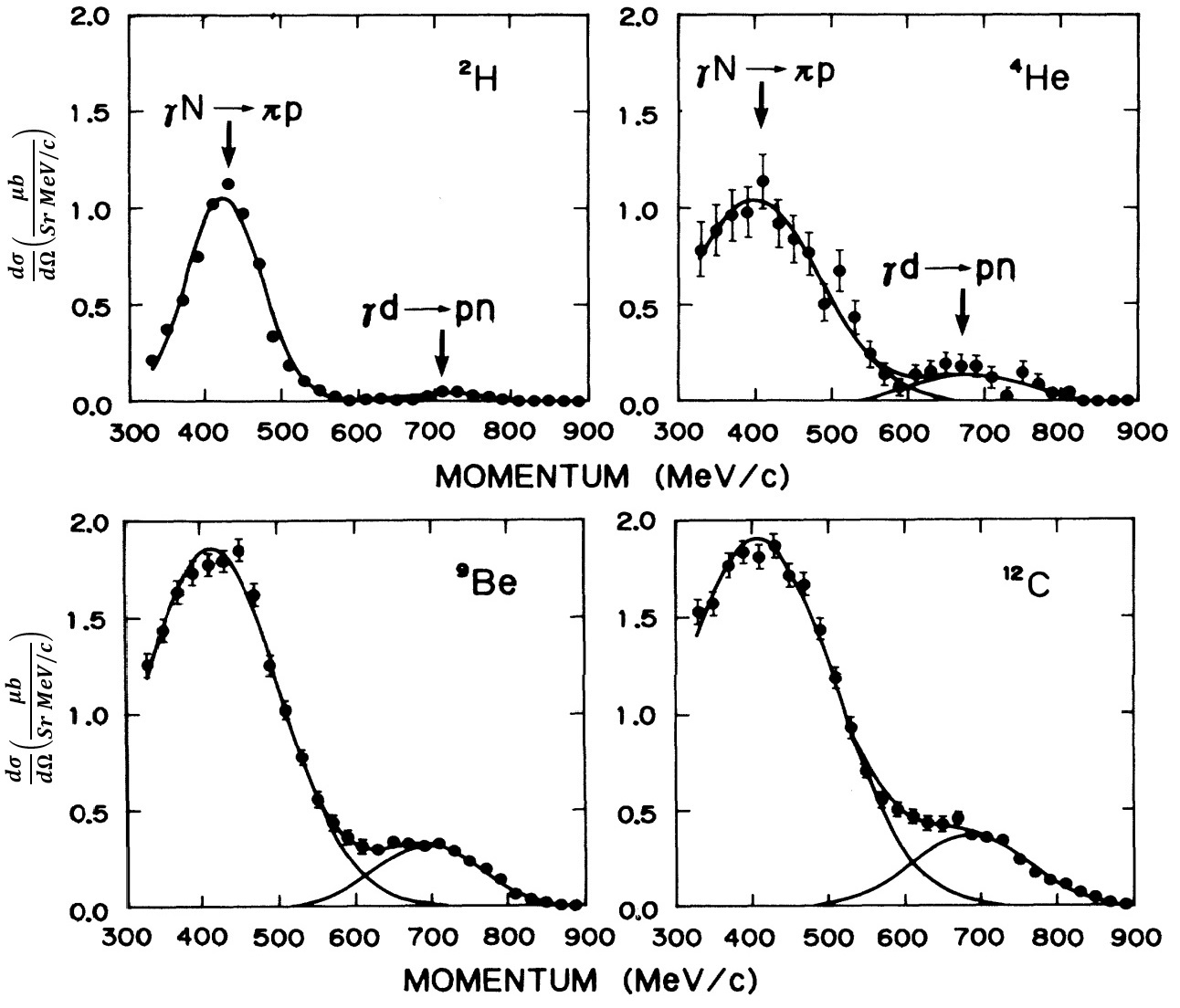}
  \caption{Differential cross-section vs momentum spectra of energetic protons at $E_\gamma$  = 357$\pm$10 MeV and 30${}^\circ$$\pm$4${}^\circ$ for different targets \cite{Homma}. Solid lines are best fits for two Gaussian functions. Vertical downward arrows for ${}^2$H target are representing the proton momenta due to the interaction of gamma with single nucleon and quasideuteron degree of freedom.}
   \label{fig-5}
\end{figure}

At higher energies, systematic investigations of the photo-disintegration of light nuclei above the pion emission threshold have now been carried out at a range of tagged photon facilities \cite{Homma, Baba2, Grabmayr, Arends}. Using photons of energy 180-420 MeV, Homma \textit{et al.} \cite{Homma} showed that energetic protons are emitted by two main mechanisms: (a) Pion photo-production from a quasi-free nucleon  $(\gamma + N \rightarrow p + \pi)$ and (b) Photodisintegration of a quasi-free two-nucleon system  $(\gamma + pN \rightarrow p + N)$. These two mechanisms can be separated kinematically and studied separately as protons ejected due to the first mechanism have significantly lower momenta due to the energy “spent” in pion production. The differential cross-section vs momentum spectra of protons for targets ${}^2$H, ${}^4$He, ${}^9$Be, ${}^{12}$C at $E_\gamma$= 357$\pm$10 MeV and observation angle 30${}^\circ \pm$4${}^\circ$ are shown in figure \ref{fig-5}. The striking similarity of momenta curves for different light nuclei demonstrates that the leading proton production mechanisms in these nuclei are same as those of  ${}^2$H. Baba \textit{et al.} \cite{Baba2} extended the validity of these conclusions for ${}^9$Be and ${}^{12}$C up to 620 MeV photon energy. The explicit nature of the proton origin from these two mechanisms has been established using coincidence measurements of protons with outgoing pions/nucleons \cite{Grabmayr, Arends}. In most of the photodisintegration events, for the quasi-free “pN”  channel, outgoing protons are accompanied by kinematically correlated neutrons, similar to the results obtained for $E_\gamma$ below pion threshold. For proton-proton emission events, the angular distribution of correlated protons is rather flat and their fraction increases rapidly with $E_\gamma$ indicating the possible role of FSI \cite{Grabmayr, Arends}. Moreover, the cross-section of protons produced through quasi-free pion production increases linearly with A and shows a $\Delta$-resonance peak. However, the cross-section for quasi-deuteron interactions scales with that of the free deuteron \cite{Homma} confirming that the validity of Levinger model extends to relatively high photon energies.\\

The success of the QDM requires rethinking the appropriate degrees of freedom in nuclei. From multiple observations, a significant fraction of nucleons inside nuclei appear to occur in quasi-deuteron formations rather than as independently moving nucleons in uncorrelated states. Further, the neutrons and protons in such quasi-deuteron configurations are expected to have higher relative momenta than the Fermi momentum ($\sim$250 MeV/c) and hence have relatively small spatial separations (less than the average NN separation of $\sim$1.4 fm). Such a set of quasi-deuterons are represented by independent pairs of unlike nucleons in the IPAM formalism discussed in section 3, where the main contributor of binding energy per nucleons arises from the two-body np interaction \cite{Gomes}.\\

\subsection{Interaction of electrons and protons with nuclei}\label{src2}

The quasi-deuteron or “spatially correlated proton-neutron pair” based photodisintegration mechanism for energetic photons in nuclei had been well-established since the inception of the QDM proposition by Levinger. However, for a long time this important signature of correlations was not studied in detail by the experimental nuclear physics community while investigating the interaction of energetic $e{}^-/p$ with nuclei. That was the case because the QDM hypothesis was not in accordance with the prevailing independent particle shell-model. The quasi-free interaction of energetic protons, having energy between 100 to 1000 MeV, has been used \cite{Jacob, Riou} to investigate nuclear shell-structure assuming quasi-free scattering $(p, p^\prime p)$. After the availability of energetic electron beams $(e, e^\prime p)$ reactions were also used extensively for the investigation of shell structure \cite{Riou}. Still, there were rather few studies which highlighted the quasi-deuteron mechanism as an important reaction mechanism in this energy region for electron beams. One example of high energy neutron production at 90${}^\circ$ was investigated for electron beam energies between 150 and 260 MeV by Eyss and Lührs at Mainz in 1973 \cite{Von}. Neutron production for 150-200 MeV was nicely accounted for by the quasi-deuteron based model by Matthews \textit{et al.} \cite{Matthews} while an additional neutron-production bump was observed for higher energy electrons due to the onset of pion-production channels. Unfortunately, no further systematic investigation was carried out for this reaction mechanism for electron or proton beams in this energy range.  \\

\begin{figure}[htbp]
  \centering
  \includegraphics[scale=0.9]{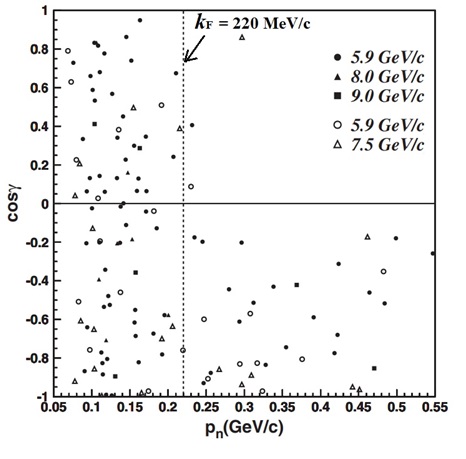}
  \caption{Scatter plot of $\bf{p_n}$ vs $\cos \gamma$ depicting the strong directional correlation between outgoing struck protons and neutrons with momentum above 220 MeV/c \cite{Pia}. }
  \label{fig-6}
\end{figure}

The simple assumption of quasi-free scattering of the energetic charged particles from protons moving independently in different nuclear shells is “spoiled” by the interaction of incoming projectiles/outgoing particles and by the presence of strong SRC of nucleons inside nuclei. The triple coincidence measurements by Malki \textit{et al.} \cite{Malki} using proton and pion beams of 5.9 GeV/c at BNL gave conclusive evidence of the strong role played by the short-range correlations in these reactions. Here, triple coincidence measurements of two elastically scattered protons (or pion-protons), along with energetic neutrons with $p_t >$320 MeV/c in the backward-hemisphere of the laboratory frame, were carried out. Interestingly, for a ${}^{12}$C target, the quasi-elastic scattering of pions/protons was accompanied by backward scattered neutrons in 40-50$\%$ of the events. These measurements were further analysed by Piasetzky \textit{et al.} \cite{Pia} by considering FSI and the expected transmission coefficients of outgoing nucleons in ${}^{12}$C. The short-ranged correlations between nucleons can be demonstrated by defining an angle $\gamma$:\\
\begin{eqnarray}
\cos \gamma = \frac{\bf{p_f} \cdot \bf{p_n}}{|\bf{p_f}| \ |\bf{p_n}|} . \label{eq3} 
\end{eqnarray}

where $\bf{p_n}$ is momentum of outgoing neutrons detected below the mid plane, while $\bf{p_f}$ is momentum of struck target proton in $(p, 2p+n)$ reaction, with $\gamma$ the angle between $\bf{p_n}$ and $\bf{p_f}$. A scatter plot of $\cos \gamma$ vs $\bf{p_n}$ is shown in figure \ref{fig-6} where it can be seen that there are no directional correlations between low momentum $(\bf{p}<$220 MeV/c) neutrons and protons, while a strong directional neutron-proton correlation is present for higher momentum struck nucleons. It was concluded that at least 74$\%$ of the time, the ejection of fast protons results in the emission of a kinematically  correlated  fast neutron. \\

Additional experiments performed at Jefferson lab using multi-GeV electron beams confirmed these results and concluded that “96 $\pm$ 22$\%$ of the $(e, e^\prime p)$ events with missing momentum above 300 MeV/c had a recoiling neutron” \cite{Subedi}. The outgoing neutron or proton may go through charge exchange reaction or FSI and may appear as nn or pp instead of a correlated np pair. The observed ratio of ${}^{12}$C$(e, e^\prime pn)$/${}^{12}$C$(e, e^\prime pp)$ indicated that at least 90$\%$ of two-body SRCs are np pairs. By considering a better FSI model and single-charge exchange (SCX) processes, it was shown that almost 94$\%$ SRCs pairs are composed of np interactions while remaining 6$\%$ are due to nn and pp interactions \cite{Duer}. These ratios are similar to those obtained earlier by McGeorge \textit{et al.} \cite{Mc} using photons, although the nucleon emission processes are somewhat different in the two energy regimes. Similar ratios have also recently been confirmed in an inverse kinematic experiment using 48 GeV ${}^{12}$C projectiles on hydrogen \cite{Patsyuk}. Moreover, SRC observations are also found to be valid for lighter \cite{Korover} as well as for heavier nuclei \cite{Subedi}. A review of experimental and theoretical progress can be found in recent work by Fomin \textit{et al.} \cite{Fomin}. \\

It has been shown that the momentum distribution of nucleons forming SRC pairs can be factorized into the CM (due to motion of SRCs inside nuclear volume) and the relative momentum distribution. The CM momentum of SRCs are found to be in range of 100-160 MeV/c for different nuclei \cite{Cohen}. On the other hand, due to much deeper NN interactions, the relative momentum distribution of nucleons inside SRCs is much higher than the expected Fermi momentum $(k_F)$. Another interesting observation is that the fraction of $(e, e^\prime pp)$/$(e, e^\prime pn)$ events increased \cite{Korover} with missing momentum from ~5$\%$ to ~12$\%$ for missing momentum increments from 500 MeV/c to 750 MeV/c. This observation is similar to the earlier findings \cite{Mac, Mc} using energetic photons at high missing energies in quasi-deuteron kinematics, indicating that some of the additional $(e, e^\prime pp)$ events at higher missing energy are originating from FSI or charge-exchange $(n,p)$ or $(p,n)$ reactions for outgoing nucleons. \\

	The observation of back-to-back CM correlations for high momentum ($>$250 MeV/c) neutrons and protons, in contrast to the absence of angular correlations for low momentum ($<$250 MeV/c) ejected nucleons, has been used to conclude an absence of NN correlations at lower energy. However, the observation of back-to-back NN correlations for low momentum np pairs originating from $(e, e^\prime np)$ or $(p, p^\prime np)$ reactions would be strongly suppressed for two reasons:
	
\begin{enumerate}
\item \textit{Low transmission coefficient:} The transmission coefficient is a strong function of the energy (or momentum) of outgoing nucleons, the type of nucleon (additional coulomb barrier for protons), and the atomic number and mass number of the nucleus concerned \cite{Blatt, CLAS1}. As an example \cite{Blatt}, only about 20$\%$ protons with $L= 0$ and energy 10 MeV are able to penetrate the nuclear surface of nucleus of $Z=$ 20 (the rest of the protons would be reflected back and may be ejected at some other angle giving incorrect information about the angular correlation of the initial interaction). For higher $Z$ nuclei and for protons ejected with higher relative angular momentum, the transmission coefficients are further decreased.

\item \textit{High interaction probability for low-energy nucleons:} For most nuclei, the GDR region is between 10-35 MeV where significant interaction cross sections for outgoing nucleons are expected. The kinetic energy of n/p with momentum $<$ 250 MeV/c overlaps with the GDR region of nuclei. Moreover, very prominent excitation function peaks of single charge exchange reactions $(p,n)$ or $(n,p)$ are also present in this energy range \cite{Kumar}.
\end{enumerate}

	For low-momentum outgoing nucleons, both of these effects take place simultaneously, hence, one cannot expect to observe strong back-to-back correlations for such nucleons. This is very similar to the difficulty faced by the experimentalists while observing the correlated neutron-proton pairs originating from $(\gamma,np)$ reactions below 50 MeV photon energy, though indirect measurements  \cite{Vey} confirmed that this reaction channel has a significant contribution toward photodisintegration processes inside the GDR region. \\

\subsection{Pion interactions with nuclei }\label{src3}

The interaction of pions with nuclei has received attention from experimentalists and theorists alike since 1950, due to their possible implications for NN interactions and nuclear structure studies \cite{Weyer}. Pions are spin-0 bosons with three charge states $(\pi^{\pm}, \pi^0)$ with rest-mass energies of 139.57 and 134.98 MeV, respectively. Pions interact with the nucleus through elastic scattering, inelastic scattering which includes charge exchange (single or double), or absorption. Of these possible reaction channels, pion absorption has been the focus of the experimental community since it is essentially a two- or multi-nucleon process due to energy-momentum conservation requirements. It can therefore be used for investigation of short-ranged spatial correlations between nucleons inside nuclei. Brueckner \textit{et al.} \cite{Brueckner} described the early pion absorption studies using a quasi-free 2N absorption model, similar to the quasi-deuteron model of Levinger \cite{Lev1}. The pion absorption cross section was assumed to be proportional to that of free deuteron (with an appropriate scale factor) and a form factor dependent on the relative motion of quasi-deuteron and spectator portion of the nucleus. \\

	The initial investigations of the pion-absorption processes were hampered by low pion beam intensities and poor energy resolution. However, early experiments were still able to demonstrate the dominance of pion absorption by quasi-deuteron, or neutron-proton SRC, in preference to proton-proton SRC \cite{Weyer, Ozaki}. The isoscaler and isovector contributions of the pion-absorbing “2N” pairs can be estimated by measuring the coincidence of kinematically opposite outgoing nucleons, though Initial State Interaction (ISI) scattering of pions pior to absorption, FSI of outgoing nucleons, and multi-nucleon absorption events may have a significant impact on various possible channels. In a stopped $\pi^-$ absorption experiment using a carbon target \cite{Ozaki}, a ratio of $\sim$5 was obtained for $nn$ emission events, originating from pion absorption on $np$ pairs by $(\pi^-,nn)$, to $np$ emission events, arising from absorption on $pp$ pairs via $(\pi^-,pp)$ or FSI. A similar result was reported by Steinacher \textit{et al.} \cite{Steinacher} using $\pi^+$ and $\pi^-$ beams of various energies on ${}^4$He nuclei. It was concluded that about 70$\%$ of pion absorption events are due to 2N mechanisms and the isoscaler $np$ channel contributes almost 20 times as much as the isovector channel. \\

The dominance of 2N quasi-deuteron absorption mechanism can also be explicitly demonstrated by measuring the ratio of cross-sections $\sigma (\pi^+,pp)$/$\sigma (\pi^-,np)$ (see review by Lee and Redwine \cite{Lee} and references therein).  A ratio of $\sim$20 was measured on ${}^3$He indicating that  2N absorption is mainly due to pion absorption by quasi-deuteron pairs. In pion absorption events, a copious deuteron production (about 10$\%$ of the proton production) was also observed. These isospin conclusions are similar to the interpretation of photon induced quasi-deuteron reactions. \\

	In addition to 2N absorption, pion-nucleus interactions also result in significant numbers of absorption events involving three or more nucleons, leading to multi-nucleon emission. The more extensive experiments at LADS at PSI and BGO at LAMPF studied their relative contributions to the total pion absorption cross-section around the $\Delta$-resonance \cite{Lee}. However, the detailed mechanisms of 3N- and multi-nucleon pion absorption processes are still not well understood \cite{Weyer, Lee}, though ISI and FSI appear to make a significant contribution. \\

\subsection{Quasi-elastic knockout of deuterons }\label{src4}

The interaction of energetic pions, photons, electrons and protons with nuclei established the presence and importance of correlated two-nucleon structures, mainly in the form of quasi-deuterons or neutron-proton SRC, inside nuclei. The presence of two-nucleon correlated structures may also enhance deuteron production in the final state, either through the quasi-elastic direct knockout of quasi-deuterons at forward angles, or by the interaction of outgoing protons/neutrons with quasi-deuterons, as reported by Imanishi \textit{et al.} \cite{Imanishi}. Deuteron production, by pickup or collision with quasi-deuterons (see figure \ref{fig-2}), could be observed even in kinematically forbidden regions for direct processes, since the initial $\pi N$ or quasi-deuteron breakup events may result in the production of secondary nucleons at large angles. \\

An interesting illustration of quasi-deuteron degree of freedom inside nuclei, through the direct knockout reaction $(N,Nd)$, was reported by Azhgirei \textit{et al.} \cite{Az} in 1958 using a 675 MeV proton beam at JINR. The momentum of outgoing scattered particles at 7.6${}^\circ$ was measured through magnetic analysis. Well-resolved deuteron peaks corresponding to elastic $(p,pd)$ scattering (shifted by the expected deuteron separation energy in these nuclei) were observed for D, Li, Be, C and O targets. Interestingly, the elastic scattered deuteron yield for these light nuclei was found to be proportional to Z. The width of the deuteron peaks provides information about the momentum distribution of quasi-deuterons, or more specifically that of the CM momentum of neutron-proton pair, inside nuclei. The kinetic energy of quasi-deuterons obtained from this study agreed with that obtained in the earlier photo-disintegration investigation \cite{Wat}. The results of this measurement were confirmed by Sutter \textit{et al.} \cite{Sutter} using a 1 GeV proton beam at BNL. From the observed average kinetic energy of these quasi-deuterons inside nuclei, Sutter \textit{et al.} concluded that these “np pairs can move rather freely through the nucleus”. Additionally, the coincident measurement of outgoing forward scattered deuterons and backscattered protons for ${}^{12}$C target revealed that resulting ${}^{10}$B nucleus is at or near its ground state. Being a loosely bound system, the elastic scattered deuterons produced in the interiors of heavy nuclei are not expected to survive for detection and free-deuteron differential cross sections proportional to $A^{1/3}$ were observed indicating that in heavy nuclei, the observed deuterons arise mainly from the nuclear surface \cite{Sutter, Az2}. The results of proton induced quasi-elastic (QE) scattering were confirmed by Riley \textit{et al.} \cite{Riley} using an 800 MeV neutron beam. Further measurements for $(p,pd)$ reaction on ${}^6$Li \cite{Albrecht} were performed with a 670 MeV proton beam and $(\pi^-,\pi^- d)$ reactions on ${}^{12}$C and ${}^{16}$O \cite{Abramov}  with 720 MeV/c pions. Recently, a 392 MeV proton beam was used to identify the spin-isospin character of neutron-proton quasi-deuteron pairs inside ${}^{16}$O through the coincidence measurements of backscattered protons and quasi-free scattered deuterons at forward angles \cite{Tara}. A strong dominance of S,T = 1,0 over S,T = 0,1 state was observed, indicating the prevalence of triplet state quasi-deuterons.\\

Systematic investigations of secondary particles ejected from a wide range of nuclei after collision of multi-GeV hadrons have been carried out since 1960. One of the first such investigations was carried out by Cocconi \textit{et al.} \cite{Cocconi} using the 25 GeV CERN Synchrotron beam at 15.9${}^\circ$ with Al and Pt targets. Unexpectedly large numbers of deuterons were observed (about 2-3$\%$ of the proton yield at that angle) which could not be accounted by either evaporation processes or by pickup processes. The production of light particles at 30${}^\circ$ for 30-GeV protons striking Al, Be, and Fe targets was investigated by Schwarzschild \textit{et al.} at BNL \cite{Sch} in which copious quantities of deuteron production were observed. The deuteron/proton ratio for Al was about 5-7$\%$ for 1-3 GeV/c momentum range of the outgoing particles. This was higher than for Be, but slightly less than for the Fe target. The deuteron/proton ratio also increased with observation angles (45${}^\circ$, 60${}^\circ$ and 90${}^\circ$). Moreover, ${}^3$H and ${}^3$He emission were also reported from Be and Pt targets with a 2.9 GeV proton beam by Pirou\'{e} \textit{et al.} \cite{Piroue}. After making allowance for the expected Fermi energy, the deuteron production for laboratory angles larger than 32${}^\circ$ should be forbidden kinematically. In contrast, copious production of deuterons was observed at 30${}^\circ$, 60${}^\circ$, and 93${}^\circ$ for Be and Pt targets. Interestingly, the ratio of total cross section for deuteron production in p-Be collisions (about 1 mb per nucleon at 2.9 GeV) and that of the reaction with free proton $p(p,d\pi^+)$ ($\sim$30 $\mu$b at the same energy) is about 33, indicating a strong quasi-deuteron presence inside nuclei. The production of ${}^3$H and ${}^3$He nuclei in these reactions can only have a nuclear origin, since their production is not expected through p-N nucleon collisions at energies of a few GeV. The production of these composite particles persisted even for proton collisions at 200 GeV energy on Be and Al targets \cite{Bozzoli}.\\

In these experiments, the production of pions and kaons arises from collisions of energetic protons with individual nucleons inside target nuclei \cite{Piroue}. For production of deuterons, tritium and ${}^3$He, Butler and Pearson \cite{Butler} suggested a coalescence model in which the incoming proton produces a cascade inside the target nucleus and the outgoing nucleons thus generated result in the formation of mass-2 and -3 composites. In this model, the deuteron formation probability is proportional to the square of the optical potential depth ($V_0$). To account for deuteron production in various nuclei at different proton energies, widely differing values of $V_0$ are required \cite{Piroue}. Moreover, sufficient cascade production quantities in light nuclei, such as Be, C and Al, would be highly improbable where almost all of the nucleons are located on or near the nuclear surface.\\

On the other hand, the copious deuteron production could be understood easily if one considers the various experimental signatures indicating the presence of quasi-deuteron, or neutron-proton SRC structures, inside nuclei as discussed above. A holistic picture can be drawn by combining the observation of recent $(p,pNN)$ measurements at the EVA spectrometer of BNL at around 27.5${}^\circ$ using proton beams of 5.9-9.0 GeV/c \cite{Pia} and mass analysis of multi-GeV proton collisions at similar angles (for example, by Pirou\'{e} \textit{et al.} \cite{Piroue} using a 2.9 GeV proton beam, or Schwarzschild \textit{et al.} \cite{Sch} employing a 30 GeV proton beam). The EVA investigation of the exclusive reaction $(p,pNN)$ demonstrated that almost all of the scattered protons  ($\sim$92$\%$) with momenta greater than 0.250 GeV/c are accompanied by a simultaneous ejection of neutrons having opposite momenta in the CM frame i.e. they were forming neutron-proton SRC just prior to their collision with the incoming energetic protons. On the other hand, owing to their experimental setting, these multi-GeV experiments are measuring high-momentum products, mainly protons and deuterons (5-7$\%$ of the protons), detected at similar angles. Hence, these experiments have measured the neutrons/protons originating (mainly) from neuton-proton quasi-deuteron or SRC breakup or direct scattered quasi-deuterons or SRC.\\

Another independent piece of evidence for possible quasi-deuteron degrees of freedom inside nuclei emerged from the copious production of deuterons in $(\gamma,d)$. Using photons from the 310 MeV Cornell synchrotron in 1953, de Wire, Silverman and Woolfe \cite{DeWire} observed deuteron to proton ratio values of about 25$\%$ to 2$\%$ at 90${}^\circ$ lab angle for Pt, Cu and C targets. A similar investigation for different targets using $E_{\gamma max}$ from 30 MeV to 90 MeV was reported by Chizhov \textit{et al.} \cite{Chizhov}, confirming a large deuteron production rate in the quasi-deuteron region. The deuteron production by photons of several hundred MeV energy ($E_{\gamma max}$ from 400 to 720 MeV) was investigated by Kihara \textit{et al.} \cite{Kihara}. A detailed investigation of deuteron emission for Be and C targets using the tagged photons of energy 360 to 600 MeV was also carried out by Baba \textit{et al.} \cite{Baba} in 1986. Photo-deuteron to photo-proton ratios of few percent to more than 20$\%$ were observed. These observations were analysed in the framework of the coalescence model of Butler and Pearson \cite{Butler} which predicted a momentum-independent deuteron to proton production ratio. An exponential ratio was observed for proton to deuteron production in all of the measurements, ruling out the Butler-Pearson model. \\

	The copious deuteron production at high scattering angles persists for pion absorption experiments \cite{Lee} below and around the $\Delta$-resonance energy. The deuteron production in the final state may account for 10$\%$ to 25$\%$ of the total pion absorption cross section in this energy domain (see \cite{Giannelli} and references therein). Similar observations have been reported for electron beams as well. Ent \textit{et al.} \cite{Ent1, Ent2} measured $(e, e^\prime d)$ reactions for ${}^4$He, ${}^6$Li and ${}^{12}$C. It was observed that most of the deuteron emission leaves the population of daughter nuclei in or around their ground state \cite{Ent1}.

\subsection{Backscattering of hadrons and sub-threshold production of energetic particles} \label{src5}

For reactions involving the interaction of an energetic probe with a single emitted nucleon, the energy transferred (W) in collision is W$\sim Q^2/(2M)$ where $Q$ is the momentum transferred to the nucleon of mass $M$. In such quasi-free knockout cases there is a momentum mismatch, which is generally absorbed by the recoiling residual nucleus. In contrast, the additional phase space available in direct absorption on two or more nucleons allows W$\ll Q^2/(2M)$. Both pion absorption and many photo-disintegration reactions are examples of processes which proceed through such a cooperative mechanism involving at least two nucleons.\\

There are two further interesting  nuclear reactions which are expected to proceed through single particle modes, if the independent particle picture is a good approximation of real nuclei, but in practice their behaviour can only be explained by assuming the cooperative behaviour of two or three nucleons (see figure \ref{fig-2}). These include:
\begin{enumerate}
\item Observation of backscattered protons/neutrons with unexpectedly large energy when nuclei are bombarded by intermediate energy (100-1000 MeV) protons \cite{Haneishi, Miake}.
\item The sub-threshold production of antiprotons and other particles in energetic nuclear collisions \cite{Knoll}. 
\end{enumerate}

Initial experiments reporting backscattered protons at intermediate proton energies were carried out at Los Alamos \cite{Frankel} and Dubna \cite{Komarov} around 1975. Frankel \textit{et al.} \cite{Frankel} impinged 600 MeV and 800 MeV protons on Be, C, Cu, Ta, Ag and Pt targets and studied the momentum distribution of backscattered protons and light fragments using magnetic fields. In these experiments, the protons are expected to collide with single nucleons in target nuclei since the reduced wavelength of impinging protons $(\lambda_p = \frac{\hbar}{p})$  is much smaller than the nucleon size. Moreover, the momentum of 600 MeV protons is much larger ($\sim$1220 MeV/c) than the expected Fermi momentum of a single nucleon ($\sim$250 MeV/c) inside these nuclei. Hence, in the context of the Independent Particle Shell Model (IPSM), one cannot expect any backscattered protons due to direct NN collisions, and scattered particles are expected in the forward hemisphere only. Still, there could be many low-energy protons/neutrons in the backward hemisphere due to particle evaporation and possible multiple scattering. However, their contribution will decrease rapidly and the number of protons/neutrons from these processes should become negligible above $\sim$50 MeV. \\

\begin{figure}[htbp]
  \centering
  \includegraphics[scale=0.9]{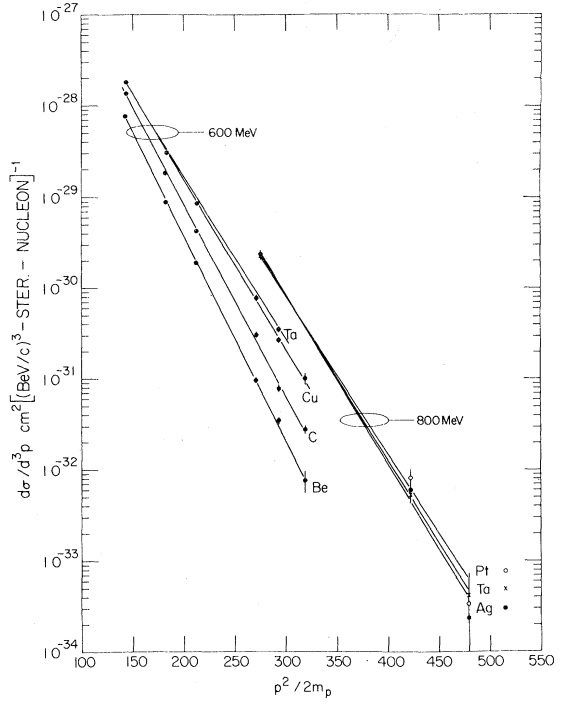}
  \caption{Differential cross-section vs kinetic energy of backscattered protons of 600 MeV and 800 MeV from different targets \cite{Frankel}. The observation points are fitted with the function $B_p\exp(-\frac{\alpha_p p^2}{ 2 m_p} )$}
  \label{fig-7}
\end{figure}

Contrary to these expectations, significant numbers of backscattered energetic protons were observed in these experiments. The differential cross-section vs kinetic energy of protons emitted around 180${}^\circ$ for protons of 600 MeV and 800 MeV is depicted in figure \ref{fig-7}.  The differential cross-section of backscattered protons was found to decrease exponentially with kinetic energy and the kinetic energy of the most energetic backscattered protons was found to extend up to 320 MeV for incident energies of 580 MeV, and up to 475 MeV for incident energies of 800 MeV. The experiment at Dubna \cite{Komarov} was carried out using 640 MeV protons with observation angles between 105${}^\circ$ and 160${}^\circ$. It also produced very similar results for C, Be, Al, Cu and Pb targets. These experiments stimulated further experimental and theoretical investigation of related phenomena \cite{Haneishi, Boal}. \\

The unexpectedly large production of energetic, kinematically forbidden, backscattered protons posed a serious challenge to the prevailing models of nuclear structure and reactions. Amado and Woloshyn \cite{Amado} calculated the observed cross-section with internal momentum of nucleons in the IPSM. But the calculated cross-section values using Gaussian momentum distributions of nucleons in the framework of single event nucleon-nucleon scattering are many orders of magnitude smaller than the measured values \cite{Amado}. A quasi-two-body scaling was proposed by Frankel \textit{et al.} \cite{Frankel2} which could account for the experimental observations using an exponential momentum distribution $exp(-k/k_0)$ with a mean momentum value $k_0\sim $90 MeV/c. \\

\begin{figure}[htbp]
  \centering
  \includegraphics[scale=0.9]{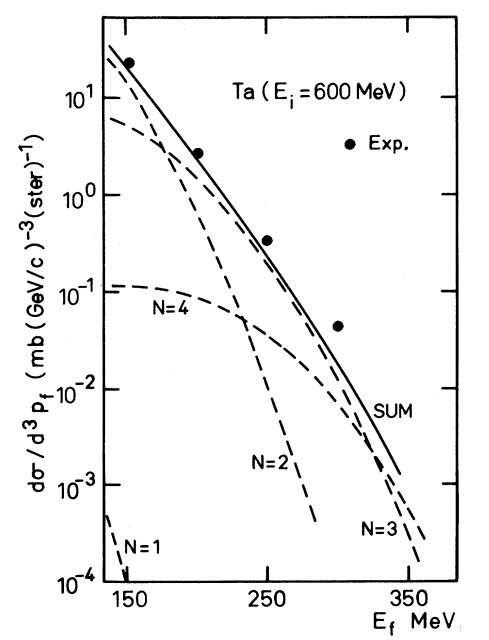}
  \caption{The contribution of different correlated clusters for differential cross-section of backscattered protons. Here, N=1 stands for single nucleons, N=2 for 2N, N=3 for 3N and N=4 for 4N clusters. This figure is taken from \cite{Fujita1}.}
  \label{fig-8}
\end{figure}

An intuitive model of few-particle “correlated clusters” was proposed by Fujita \cite{Fujita1} and further elaborated by Fujita and Hüfner \cite{Fujita2} which showed that the direct scattering from “quasi-deuterons” or 2N correlated clusters and from 3N correlated clusters may account for the observations in a consistent way. A direct collision of incoming protons with two- and three-nucleon “correlated clusters” can transfer large momentum, leading to observed backscattered protons with high-energy. The estimated backscattered cross-section contributions from individual nucleon, 2N, 3N, 4N clusters is depicted in figure \ref{fig-8}.  

Knoll and Shyam \cite{Knoll} showed that “correlated clusters” arising from statistical processes may account for the observed backscattering of protons. Stelte and Wiener \cite{Stelte} extended this approach by examining the relation between backscattered proton cross-sections, for incident beams up to 400 GeV, and thermodynamic energy-momentum “Hot Spots” inside nuclei. They demonstrated an exponentially falling dependence of the backscattering cross section on the kinetic energy of the emitted proton. A detailed review of experimental and theoretical work up to 1981 by Frankfurt and Strikman on this topic can be found in \cite{Frankfurt2}.\\

It has become apparent from this work that few-nucleon correlated structures inside nuclei may be responsible for the observed backscattered proton flux in the above-mentioned observations. This has motivated searches for direct evidence of the role of correlated structures through coincident measurements of backscattered protons with additional particles scattered in the forward hemisphere. One excellent study, carried out by Miake \textit{et al.} in 1985, used an 800 MeV proton beam at Berkeley \cite{Miake}. Energetic protons with momentum greater than 350 MeV/c were detected at $\sim$118${}^\circ$ in coincidence with particles ejected between 15${}^\circ$ and 100${}^\circ$. The dominance of proton backscattering from “correlated clusters” was verified by this measurement. Further, the momentum distribution of the np quasi-deuteron “correlated clusters” was determined. The observed momentum spread of “correlated clusters” of 85$\pm$15 MeV/c is not far from earlier measurements using photons \cite{Wat} and later experiments with energetic proton or electron beams \cite{Cohen}. This measurement was also in accordance with the earlier works of Azhgirei \textit{et al.} and Sutter \textit{et al.} and with much more recent RIKEN measurements \cite{Tara} using 392 MeV protons on an ${}^{16}$O target.  \\

The sub-threshold production of particles is understood as particle production in particle-nucleus or nucleus-nucleus collisions in which the available CM energy per nucleon is below the particle production threshold for free NN collisions. The sub-threshold production of anti-protons $(\bar{p})$, pions, kaons and other sub-atomic particles has been investigated since the mid-1950s. \\

Using a proton beam on a Cu target, Chamberlain \textit{et al.} \cite{Chamberlain} measured the production of anti-protons with a momentum of 1.19 GeV. Anti-proton production was observed at 4.9 GeV (lab energy), below the free NN threshold of 5.6 GeV. Dorfan \textit{et al.} \cite{Dorfan}  reported the $\bar{p}$ production for reaction $p + Cu \rightarrow \bar{p} + $residue, for proton kinetic energies between 6.1 GeV and 2.88 GeV. For these energy values, the incident momenta $\gg \hbar/R$, where R is the p or n radii and incident proton is expected to interact with individual nucleons of target nucleus. The $\bar{p}$ production probability was computed assuming  nucleonic kinetic distribution $g(t)$ as Gaussian type. Using “$t$” as energy of target nucleon, Gaussian type kinetic energy distribution is:\\
\begin{eqnarray}
g_g (t) = \exp[-(t/t_0)^2] \label{eq4} 
\end{eqnarray}\\
 or alternatively a Hulth\'{e}n type distribution:\\
 \begin{eqnarray}
g_h (t) \varpropto \left( \frac{1}{2t + \alpha^2} - \frac{1}{2t + \beta^2} \right)^2 \label{eq5} 
\end{eqnarray}\\

It was observed that the Gaussian $\bar{p}$ production probability could be fitted only for the exceptionally high value of mean energy $t_0 =$ 60 MeV instead of the lower value $t_0=$ 20 MeV obtained from low energy experiments. \\

	The differential cross-section for the subthreshold $\bar{p}$ production was measured by Pirou\'{e}  \cite{Piroue} using 2.9 GeV protons on Be and Pt targets. Confirming the previous study by Dorfan \textit{et al.}, the $\bar{p}$ production widely disagreed with Gaussian energy distribution of target nucleons with $t_0 =$ 20 MeV. A Hulth\'{e}n type momentum distribution $(f(p) d^3 p \propto d^3p / [(p^2 + \alpha^2)(p^2 + \beta^2)]$ with cutoff momenta $p_{max} = $1 GeV/c, $\alpha = $0.047 and $\beta = 7\alpha$ could be used for a approximate reproduction of measured values. However, the measured cross-section could be reproduced nicely for a momentum distribution having extended tail;\\
\begin{eqnarray}
f(p) d^3 p \propto \frac{d^3p}{1 + e^{(p-p_0)/\lambda}}  \label{eq6} 
\end{eqnarray}\\

\begin{figure}[htbp]
  \centering
  \includegraphics[scale=0.9]{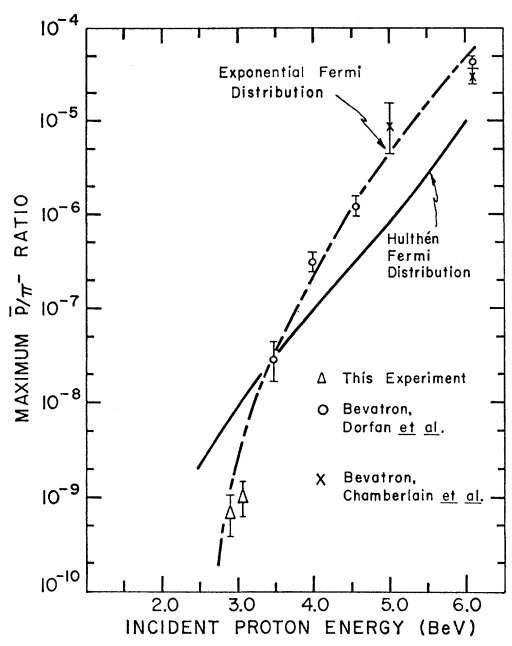}
  \caption{Plot of $\bar{p}/\pi^-$ ratio vs incident proton energy (GeV). This figure is taken from \cite{Piroue}}
  \label{fig-9}
\end{figure}

The $p_0$ value of 0.10 GeV has been suggested which results in mean nucleonic energy equal to 20 MeV. The measured and calculated results, expressed in $\bar{p}/\pi^-$ ratio are depicted in figure \ref{fig-9}.\\

Since the subthreshold $\bar{p}$ production is dictated by the high momentum tail in the nucleonic momentum distribution, a single step scattering of incoming projectile nucleon by target nucleons with momentum distribution $F(k)=(\frac{1}{k})\exp(-k/k_0)$ can be used to fit the $\bar{p}$ cross-sections for different incident proton energy values \cite{Frankel3}. This is the same function previously used by Frankel \textit{et al.} \cite{Frankel2} to account for backscattered particles as mentioned previously.\\

\begin{figure}[htbp]
  \centering
  \includegraphics[scale=0.7]{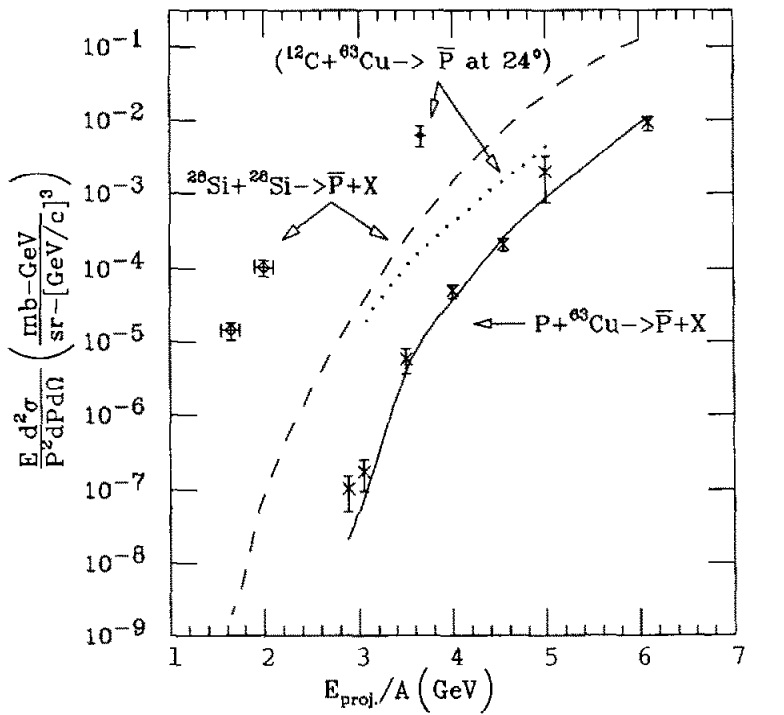}
  \caption{Comparison of sub-threshold $\bar{p}$ production in $p + Cu$ and $Si + Si$ reactions \cite{Shor}. The calculated cross-section values (solid line) obtained by assuming high momentum component for nucleons (parameterized using backscattered proton data) agree well with measured sub-threshold $\bar{p}$ production in p + Cu collisions. However, similar calculations for Si + Si (dashed line) and C + Cu (dotted line) underestimate the experimental cross-section values by large factors.}
  \label{fig-10}
\end{figure}

Several objections have been raised for simply using the high momentum tail of the nucleon distribution as a plausible explanation for the sub-threshold particle production: 

\begin{enumerate}
\item The single particle Fermi gas model results in a Gaussian-type momentum distribution for the nucleons with steep fall in high momentum component. The high-momentum exponential-tail in the momentum distribution arises from empirical fitting of $\bar{p}$ production cross-section only; 

\item Coincident measurements between backward scattered energetic protons and forward scattered particles by Miake \textit{et al.} \cite{Miake} demonstrated that the main contribution to backward energetic protons arises from the presence of 2N or heavier clusters inside nuclei and hence sub-threshold production of $\bar{p}$ may also arise from the presence of short-range correlated nucleons; 

\item Shor \textit{et al.} \cite{Shor} measured the sub-threshold $\bar{p}$ production for p-nucleus and in nucleus-nucleus collisions. The sub-threshold $\bar{p}$ and $K^+$  production in p-nucleus collisions can be accounted for very well using the internal momentum distribution obtained from backscattered proton data. However, the similar approach underestimated the sub-threshold $\bar{p}$ and  $K^+$  production cross-sections in nucleus-nucleus collisions by large factors. In these calculations, the internal momentum distribution for projectile nucleons is used together with parameters obtained from p-nucleus collisions. A comparison of calculated and measured results is shown in figure \ref{fig-10}.
\end{enumerate}

\begin{figure}[htbp]
  \centering
  \includegraphics[scale=0.8]{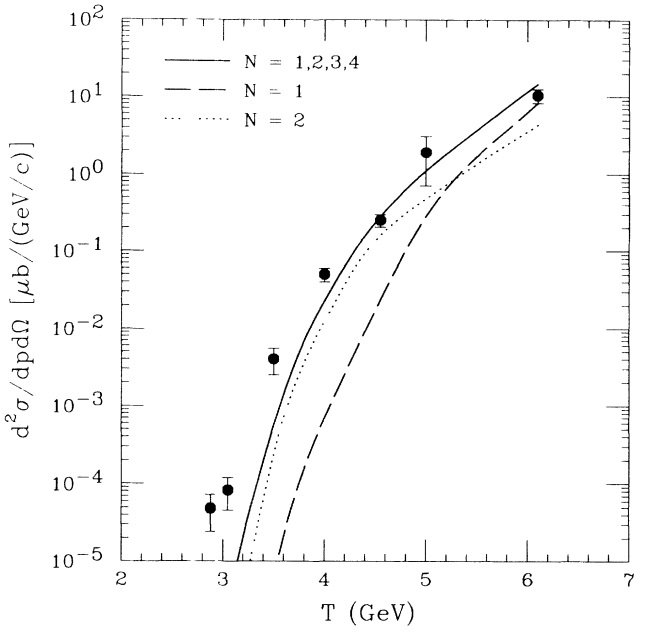}
  \caption{$\bar{p}$ production cross-section vs proton kinetic energy in $p + Cu \rightarrow \bar{p} + X$. Long dashed line depicts the $\bar{p}$ production cross-section contribution from individual nucleons while the dotted line indicates the contribution from two-nucleon clusters \cite{Dani}. }
  \label{fig-11}
\end{figure}

The idea of short-ranged correlated nucleons was introduced by Frankfurt and Strikman \cite{Frankfurt2} for possible coherent interaction of projectile with more than one nucleon. In addition to the Fermi motion of nucleons, Danielewicz \cite{Dani} included multi-particle interactions to calculate the cross-section of backscattered protons and sub-threshold $\bar{p}$ production. The production cross-section of $\bar{p}$ is expanded in the number of nucleons involved; $R = R_2 + R_3 + \cdot \cdot$. The $\bar{p}$ production cross-section for $p+Cu \rightarrow \bar{p} + X$ is depicted in figure \ref{fig-11}. Clearly, the $\bar{p}$  production in sub-threshold region cannot be accounted by NN interactions and two-nucleon clusters are required to account for the observed cross-sections. \\

\begin{figure}[htbp]
  \centering
  \includegraphics[scale=0.7]{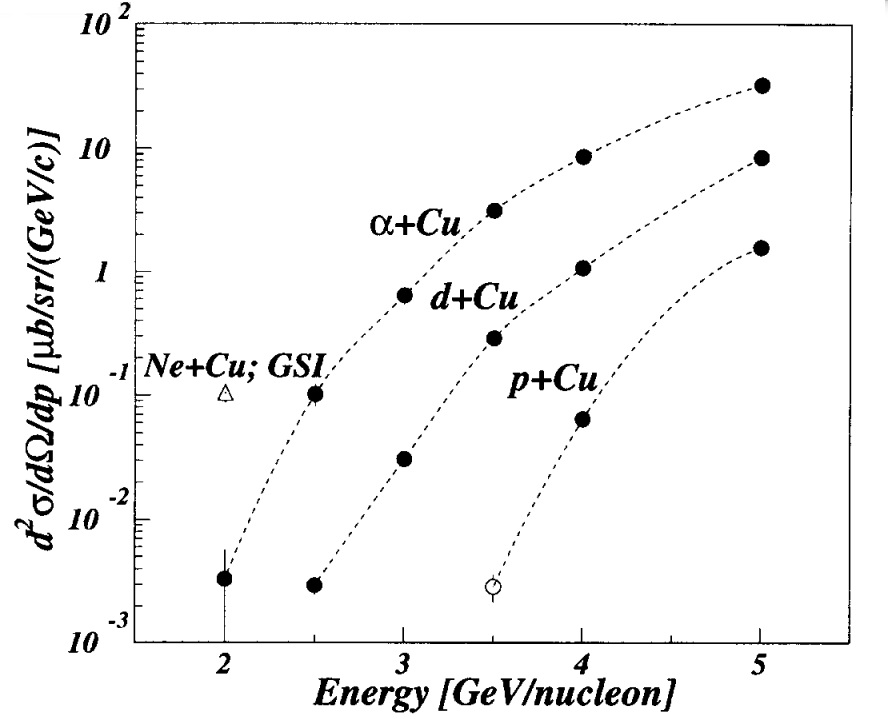}
  \caption{$\bar{p}$ production cross-section with 1.5 GeV/c for p + Cu, d + Cu and $\alpha$ + Cu reactions. An earlier measurement of $\bar{p}$ cross-section in Ne + Cu is also shown.  }
  \label{fig-12}
\end{figure}

Further understanding of the sub-threshold $\bar{p}$ production mechanism was achieved through  systematic investigations of reactions like p + Nucleus, d + Nucleus and He + Nucleus by Sugaya \textit{et al.} \cite{Sugaya}. These studies measured the $\bar{p}$ production cross-sections using proton, deuteron and alpha beam for targets C, Al, Cu and Pb. The projectile energy ranged between 2 to 5 GeV/A. At an incident energy of 3.65 GeV/A, two orders of larger cross-section were observed for the $\bar{p}$ production in the d + Cu reaction compared to p + Cu, as shown in figure \ref{fig-12}. Such a huge increment in the  $\bar{p}$ production cross-section cannot be explained by the internal momentum of nucleons in the loosely bound deuteron. This investigation was complemented by measurements by M{\"u}ller and Komarov \cite{Muller} for p + C, d + C and $\alpha$ + C reactions. The huge enhancement in $\bar{p}$ production cross-sections for the d + C and $\alpha$ + C reactions  are observed. It was concluded that the presence of few nucleon clusters is only viable explanation for these observations.\\

A strong dependence on nuclear media was also observed for the production of other sub-atomic particles like kaons and pions around and below their kinematic threshold for free NN collisions.  A detailed review of sub-threshold production of pions in heavy ion collisions was carried out by Braun-Munzinger and Stachel \cite{Braun}. The sub-threshold kaon production for ${}^{36}$Ar projectile with 92 MeV/A energy was measured for ${}^{12}$C, ${}^{nat}$Ti and ${}^{181}$Ta targets \cite{Legrain}. Assuming validity of the Fermi-momentum distribution, Bertsch \cite{Bertsch} showed that the pion production cannot take place in nucleus-nucleus collisions below 50 MeV/A. On the other hand, numerous studies have pointed out that the pion production can take place at much lower energy values and the absolute threshold for these reactions could be as low as about 10-20 MeV/A (depending on the masses of the colliding nuclei). Similar observations were reported \cite{Noll} for neutral pion $\pi^0$ production in sub-threshold region using 64-84 MeV/A ${}^{12}$C projectiles on a range of different target nuclei from A=12 to A=238. The production of pions in “kinematically forbidden” regions appears due to the co-operative action of multiple nucleons in the target and the projectile. A range of models have been proposed which assume different degrees of collectiveness inside nuclei \cite{Frankfurt2, Vasak, Knoll2}. The kinematic behaviour of pion production indicated that it is dominated by a cooperative mechanism with a coherent interaction of many nucleons from two colliding nuclei. \\

\subsection{Neutrino interactions with nuclei}

The detailed understanding of neutrino-nucleus interaction is paramount for the success of next-generation neutrino oscillation experiments like DUNE and Hyper-Kamiokande \cite{Katori, Acciarri}. Apart from reducing the systematic errors in neutrino-oscillation studies, such investigations can be used to illuminate the short-range correlation inside nuclei through the two-body $(2p$-$2h)$ excitation mechanism, in which the neutrino interacts with a correlated nucleon pair. Until recently, the focus of neutrino interaction measurements have been the charged current quasi-elastic (CCQE) interactions of $\nu_l/\bar{\nu}_l$ where $l$ denotes lepton flavour. The CCQE with bound nucleons progress through production of the corresponding lepton flavour; \\
\begin{eqnarray}
\nu_l + n \rightarrow l^- + p, \ \ \  \bar{\nu}_l + p \rightarrow l^+ +n \label{eq7} 
\end{eqnarray}

In general, the measured CCQE events with no pion in final state do not distinguish the genuine CCQE contribution from that arising due to two other important channels. One is the pion production at interaction vertex and subsequent absorption of those pions inside the nuclear medium. As discussed in previous sections, pions are strongly interacting particles and pion absorption probability would be very high even for light nuclei like ${}^{12}$C. Further, the contribution of $(2p$-$2h)$ events is not distinguished from that of the genuine CCQE events originating from $\nu_l/\bar{\nu}_l$ scattering from single nucleons. These limitations persist in most of the present and future neutrino detectors using C, O or heavier target atoms like Ar in the Deep Underground Neutrino Experiment (DUNE) and the Short Baseline Neutrino (SBN) program at Fermilab. Due to these two additional contributions, the measured neutrino-nucleus cross-sections values would likely be higher than those calculated assuming validity of the impulse approximation for Fermi gas model of nuclei. \\

\begin{figure}[htbp]
  \centering
  \includegraphics[scale=0.6]{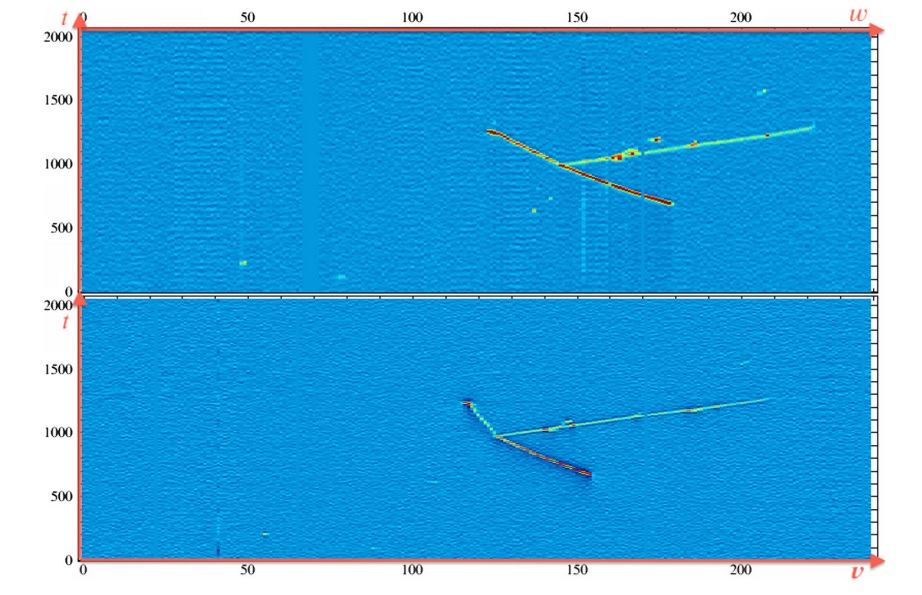}
  \caption{2D depiction of a $2p-2h$ “hammer event” with back-to-back-protons and forward moving $\mu^-$. $\omega$, $t$ and $v$ are transformed TPC wire-plane coordinates from the TPC detector \cite{Acciarri}. }
  \label{fig-13}
\end{figure}

Recently, neutrino interaction events with two-body SRCs \cite{Acciarri} have been observed directly in ArgoNeuT detector in the NuMI (Neutrinos at the Main Injector) low energy beam line of Fermi Lab. The ArgoNeuT detector is designed to detect the charged particle tracks in liquid Ar. The CCQE interaction of $\nu_\mu$ with average energy $〈E_\nu 〉\sim$4 GeV with 2N SRCs results in characteristic “hammer events” with two back-to-back protons and a forward going $\mu^-$. A 2D view of one such hammer event is shown in figure \ref{fig-13}.\\

The observed discrepancy between measured neutrino cross-section and theoretical predictions based on the relativistic Fermi gas model for nuclei can be resolved by inclusion of multi-particle mechanisms \cite{Katori, Martini, Nieves, Bonus}. Martini \textit{et al.} \cite{Martini2} computed the contribution of various channels for electron-neutrino cross-section for T2K experiments and found that a significant contribution from multi-nucleon $(np$-$nh)$ scattering channel is required to account the experimental results. Since, the contribution of these events may bias the neutrino-nucleus cross-section and energy construction, intense theoretical activities aimed at modelling these interactions have been initiated \cite{Katori, Nieves}. Unfortunately, neutrino detectors are not designed to distinguish $2p$ events from single particle charge current quasi-elastic events and only the sum of these events are measured. Phenomenological models \cite{Nieves, Bonus} are being used to predict neutrino-nucleus interaction, with the effect of short-range correlations, on top of Fermi-motion of nucleons, being implemented in Monte-Carlo event generators \cite{Campbell}.\\ 

\subsection{EMC effect}

Lepton scattering from nuclei provides important information about the quark distribution inside bound nucleons and inclusive deep inelastic  $d^2\sigma/dxd(Q^2)$ measurements can be used to estimate the structure functions $F{}^A_1$ and $F{}_2^A$ which depend on the properties of the quark distribution. In the Bjorken limit, $F{}^A_1$ and $F{}_2^A$ become independent of the four-momentum transferred square i.e. $Q^2$, and $F{}^A_1$ and $F{}_2^A$ are given by:
\begin{eqnarray}
 F_1(x) = \frac{1}{2}\sum_q e{}^2_q q(x), \ \ \ \ F_2(x) = 2x F_1(x)      \label{eq8} 
\end{eqnarray}

\begin{figure}[htbp]
  \centering
  \includegraphics[scale=0.6]{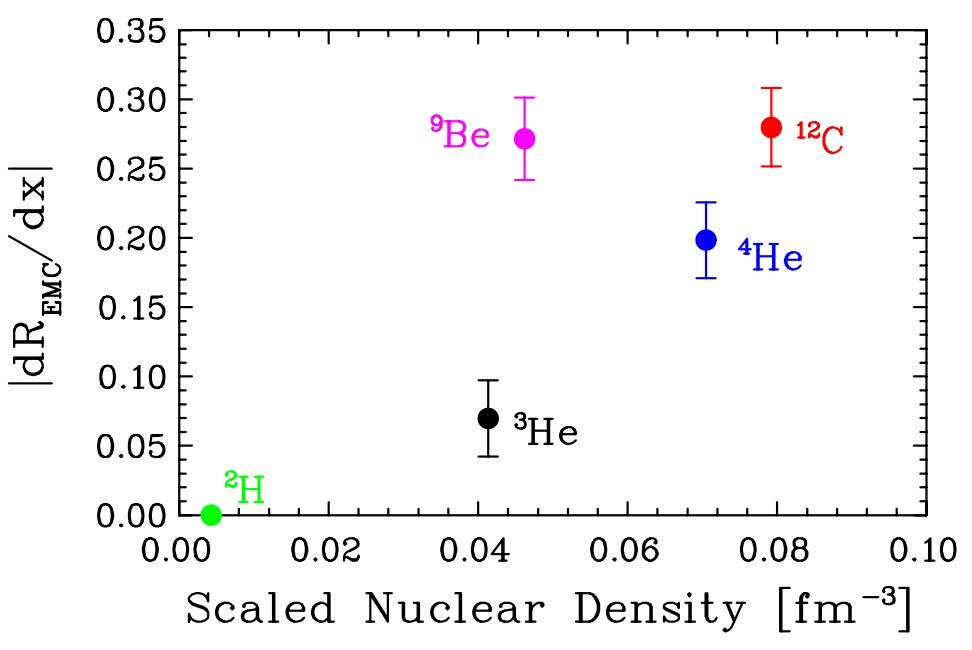}
  \caption{Plot of $|dR_{EMC}/dx|$ vs nuclear density for some of the light nuclei. This figure is taken from \cite{Seely}. }
  \label{fig-14}
\end{figure}

Assuming that the transverse and longitudinal cross-section ratio is independent of A, the cross-section would be proportional to $F{}_2^A$. Hence, we have; $(F{}_2^A/A)/(F{}_2^D/2) = (\sigma_A/A)/(\sigma_D/2)$. The variation of $(\sigma_A/A)/(\sigma_D/2)$ with Bjorken parameter $x$  provides information about the quark distribution modification for bound nucleons w.r.t. the loosely bound deuterium nucleus. Since the energy scale of virtual photon-quark interaction for 0.3$\leq x \leq$0.7 is much higher compared to the few MeV binding energy of nucleons inside nuclei, it is expected that $R_{EMC} = (\sigma_A/A)/(\sigma_D/2)$ $\sim$ 1. However, the European Muon Collaboration \cite{EMC} observed that the Deep Inelastic Scattering (DIS) structure functions of the bound nucleon in Fe are significantly different from those in deuterium for the above specified $x$ region. These observations, termed the “EMC effect”, were confirmed by subsequent measurements at SLAC for ${}^4$He, Be, C, Al, Ca, Fe and Ag targets \cite{Gomes, Sargsian}. In the literature, the EMC effect is quantified by the gradient $|dR_{EMC}/dx|$ for 0.35$\leq x \leq$0.7.  These experiments indicated that EMC effect is scaling with $log[A]$ and was proportional to the nuclear density \cite{Sargsian}. However, more detailed investigations involving light nuclei did not support these conclusions and indicated that the EMC effect may be dependent on the local environment of the nucleon \cite{Seely}. A plot of  the EMC ratio $|dR_{EMC}/dx|$ vs scaled nuclear density is depicted in figure \ref{fig-14}.

These results cannot be explained by the usual independent particle approach where nucleons are moving in a mean field generated by all other nucleons. On the other hand, a natural explanation of the EMC effect may be found in terms of short-ranged correlation of nucleons inside nuclei \cite{Weinstein, Arrington}. The strong NN interaction would invariably lead to the significant wavefunction or quark-structure overlap of nucleons inside nuclei. This would lead to modification of quark structure function of nucleons inside nuclei compared to free or loosely bound nucleons. The relationship between the EMC effect and the SRC probability can be visualised by exploring the possible linear correlation between these effects, by plotting $|dR_{EMC}/dx|$ vs the $a_2 (A/d)$ parameter for various nuclei \cite{Hen}. Here, $a_2 (A/d)$ denotes the  probability of finding a nucleon in a 2N cluster for a given nucleus compared to that of for a deuteron. The $a_2 (A/d)$ parameter is defined as:
\begin{eqnarray}
 a_2(A) = \frac{2}{A} \frac{\sigma_A(x, Q^2)}{\sigma_d(x, Q^2)}      \label{eq9} 
\end{eqnarray}

\begin{figure}[htbp]
  \centering
  \includegraphics[scale=0.6]{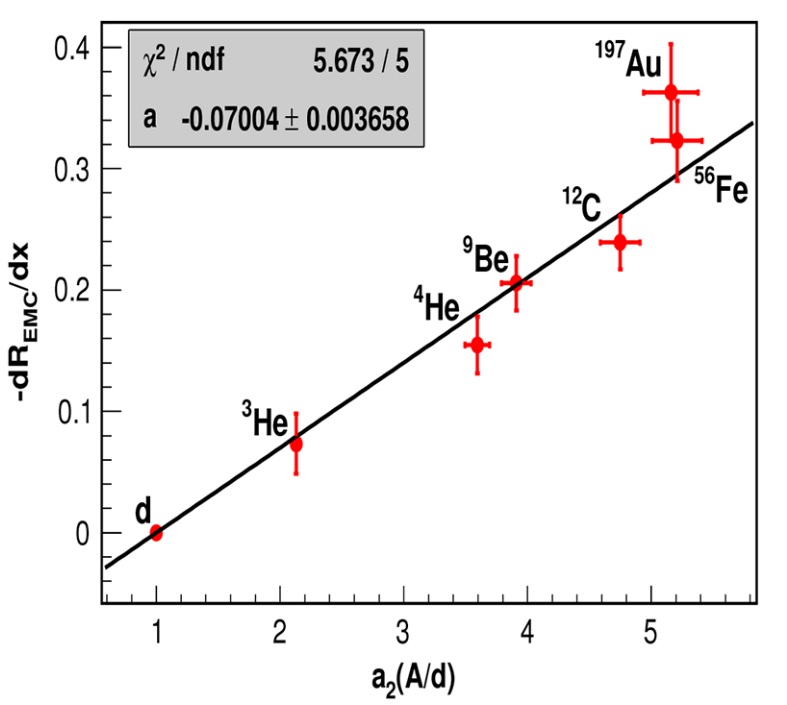}
  \caption{Plot of  $|dR_{EMC}/dx|$ vs $a_2 (A/d)$ for various nuclei. Figure taken from \cite{Hen}. }
  \label{fig-15}
\end{figure}

Here, $\sigma_A (x,Q^2 )/A$ denotes the inclusive per nucleon cross-section for $(e,e^\prime)$ for a given nucleus for Bjorken parameter $x$ and four momentum squared $Q^2$. For $x>$1, only 2N and 3N cluster would contribute toward inclusive $(e,e^\prime)$ cross-section. Neglecting the contribution of 3N clusters, one would expect to get a plateau for $a_2 (A)$ in the region 1$<x<$2. However, due to smearing of 2N cluster momentum, the corresponding flat $a_2 (A)$ region is observed for 1.5$<x\leq$2. A plot of $|dR_{EMC}/dx|$ vs  $a_2 (A/d)$ parameter is shown in figure \ref{fig-15}. It can be observed that a good correlation exists between the EMC effect quantified by the gradient $|dR_{EMC}/dx|$ and the 2N cluster probability parameter $a_2 (A/d)$ \cite{Hen}. \\

\begin{figure}[htbp]
  \centering
  \includegraphics[scale=0.6]{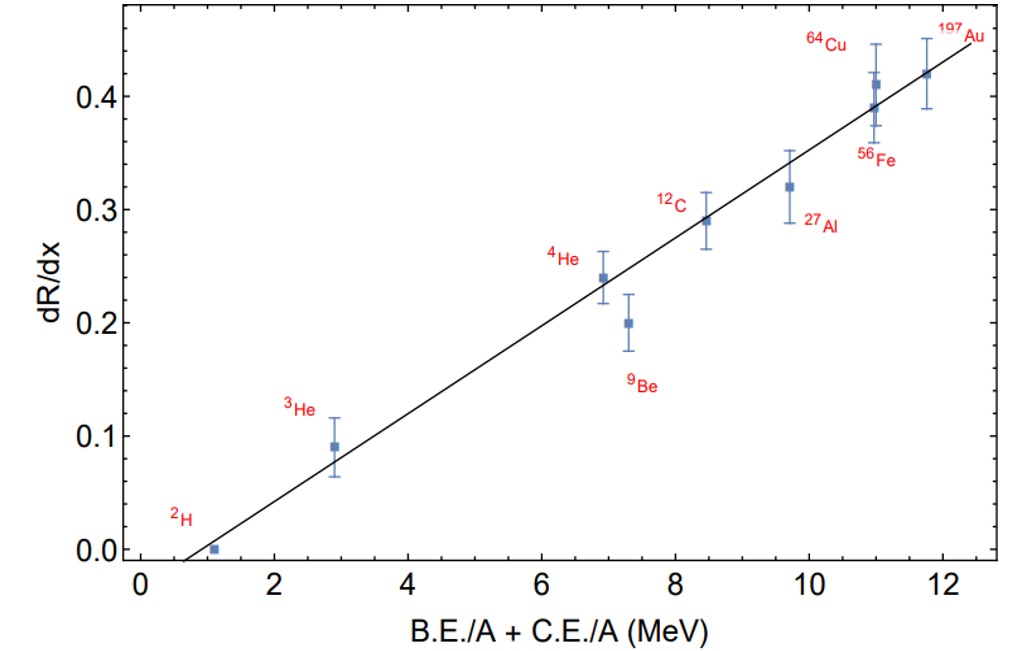}
  \caption{$dR/dx$ vs corrected B.E./A (sum of observed B.E./A and estimated Coulomb energy per nucleon calculated using liquid drop model with $a_c$ = 0.717 MeV) \cite{Ranjeet}. }
  \label{fig-16}
\end{figure}

Another way to visualize the effect of strong interactions on EMC effect is that the nuclear binding energy per nucleon (observed B.E./A corrected for appropriate Coulomb energy) arises mainly from the interaction of nearest-neighbours inside nuclei. These NN interactions may modify the quark structure function of interacting nucleons leading to a correlation between corrected nuclear B.E./A  and the $|dR_{EMC}/dx|$ ratio. Interestingly, as shown in figure \ref{fig-16}, a linear correlation between these two quantities has been observed \cite{Ranjeet}. Similar correlations were independently reported by Wang and Chen \cite{Wang}.\\

\section{Theoretical frameworks for SRCs}

As discussed in last section, there exist diverse sets of experimental evidence advocating  the existence of strong N-N interactions inside nuclei which have inspired many nuclear physicists to develop suitable formalisms for nuclear structure to account for these observations. Many of these models have already been highlighted above in discussing the experimental observations. Some additional prominent approaches encompassing the short-range nucleonic correlations are discussed here.\\

The earliest successful employment of short-range NN correlations for understanding the capture process of $\pi^-$ meson in nuclei is due to Tamor \cite{Tamor}. He assumed that the $\pi^-$ mesons are captured by a very close two-nucleon system while rest of the nucleons act as spectators. Heidmann \cite{Heid} successfully employed the two-nucleon model for pick-up reactions involving 90 MeV neutrons. The success of quasi-deuteron model of Levinger for photo-disintegration processes, along with other models like 2N pion-absorption mechanisms, further cemented confidence that there is a strong spatial correlation between neutrons and protons, and that many body systems have not fully averaged out these interactions inside nuclei. \\

However, the success of the nuclear shell model was interpreted in terms of the existence of long mean free paths for the constituents and the independent particle motion was considered as the absence of strong two-body interactions inside nuclei. Brueckner, Levinson and Mahmoud \cite{Brueckner2} developed an alternative theoretical approach by modifying “Independent particle wave function” $\Phi_0$ by a transformation function or model operator F which has the effect of introducing NN correlations. A brief review of experimental evidence for strong NN correlations till 1955 along with a possible explanation through their NN correlation model is given by Brueckner, Eden and Francis \cite{Brueckner3}. The Brueckner approach was further expanded by Bethe \cite{Bethe}. \\

Considering an approximate NN “Serber” central force represented by a finite potential-well with repulsive core (neglecting tensor force and assuming that the NN interaction inside nuclei is that of the isolated pair), Gomes, Walecka and Weisskopf \cite{Gomes} (GWW) investigated the properties of nuclear matter for N=Z. In this Independent PAir Model (IPAM), the interaction between nearest nucleon pairs was considered and the resulting potential and kinetic energy terms were calculated using the Beth-Goldstone (B-G) equation. The Fourier components of wavefunctions belonging to the occupied levels up to the Fermi energy level, except those occupied by pair forming nucleons, were rejected using projection operator $Q{}_{\alpha \beta}^F$. The B-G equation, thus formed, can be separated according to the relative angular momentum quantum number. The NN pairs in relative angular quantum number L = 0 will be due to neutron and proton only, since pairs of $nn$ and $pp$ are not allowed due to requirements of anti-symmetry. Only the S-part (L = 0) of the $\Psi_{BG}$ wavefunction would be physically realizable since the Serber force excludes odd-states and dominating centrifugal terms for the small relative NN separations for D-state interactions. The calculated binding energy per nucleon contains three terms, kinetic energy of the nucleon (Fermi-gas) in nuclei, the relative kinetic energy of motion of nucleons and the interaction energy of the pair (both of these are due to NN interaction). The calculated binding energy per nucleon from this approach is only about half of the experimental volume energy value. However, the error is only about 12-14$\%$ of the attractive potential energy and may be due to neglected tensor forces and partially may be due to the stronger NN interaction between nucleons inside the nuclear medium. Analogues to the SRC observations, the IPAM also allows two types of motion for the nucleons inside nuclei, the relative motion between nucleons due to the NN interaction and the Fermi-gas motion inside nuclei.\\

The qualitative results of GWW were confirmed by Forest \textit{et al.} \cite{Forest} through microscopic calculations using Reid, Paris, Urbana and Argonne v18 potentials for the NN interaction. The short-range structures for ${}^2$H, ${}^{3,4}$He, ${}^{6,7}$Li and ${}^{16}$O were examined by calculating two-nucleon density distributions for specified spin-isospin quantum numbers. The two nucleon T,S = 0,1 density distributions in these nuclei result in a toroidal shape for M${}_S$=0 and in dumbbell shapes for M${}_S$ = $\pm$1. Moreover, these shapes were found to be similar to those of the deuteron except for a scale factor. Since these spatial SRCs of nucleons cannot be represented in the independent particle shell model framework, these have motivated alternative frameworks like unitary correlation operator method (UCOM) \cite{Roth} which includes these correlations as prior condition. \\ 

         Apart from the anti-symmetry requirement leading to additional possibility of np pair interaction in L= 0 relative angular momentum states, there is another argument for the stronger np interaction compared to $pp$ and $nn$. The electromagnetic potential energy of two protons would increase rapidly if there is significant overlapping of protons wave functions. Similarly, the overlapping of two neutron wave functions would be hindered by the electromagnetic interaction since there is net negative charge density in outer region of neutron \cite{Lorce}. \\
         
On the other hand, an $np$ interaction will be supported by the electromagnetic interaction of positive charge distribution in the proton and the outer negative charge density in the neutron. Since the charge polarization of the neutron-proton system due to such an np bonding inside nuclei has never been measured, it is difficult to estimate the fraction of electromagnetic interaction energy contribution in np binding energy. However, it could be a significant fraction of the NN binding energy, since the minimum of the NN interaction is at a separation of $\sim$ 0.8-0.9 fm, which is similar to the size of nucleons. \\

By considering a two-component Fermi-gas inside nuclei, Weiss \textit{et al.} \cite{Weiss} extended Tan’s work \cite{Tan} on the zero-range model for ultra-cold atomic system to the nuclear medium. A nuclear wavefunction factorisation similar to GWW is assumed where strongly interacting two-body wavefunctions with relative s-wave contribution is considered \cite{Weiss2}. The contact formalism is widely used to analyse the recent NN correlation measurements at Jefferson Lab and Brookhaven National Laboratory \cite{CLAS, Cruz}.\\

\section{Future presepectives}

The past few decades have significantly extended our understanding of short-range NN correlations through a wide range of experimental and theoretical work as discussed in the previous two sections. However, many significant issues pertaining to short-range NN correlations remain unresolved or require further detailed investigation. These include the relation between the EMC effect and 2- and 3-nucleon SRC structures, variation of NN correlations with nuclear density and momentum, the role and extent of 3N correlations (especially their contribution to nuclear binding energies), and modification of NN correlations with isospin asymmetry in nuclei. The answers to these questions may pave the way for the QCD-based microscopic description of nuclei, integration of NN correlations into nuclear structure models, and a deeper understanding of the interaction of elementary and composite particles such as neutrinos and mesons with nuclei. Further, SRC physics has important ramifications for the equation of state for extended nuclear matter exemplified by neutron stars. The presence of short-range correlations between protons (constituting only few percentage of neutron star mass) and neutrons would lead to the much higher Fermi momentum to protons which in turn would affect the thermal evolution of these stars \cite{Sales}. The SRCs modified single nucleon momentum have been incorporated in relativistic mean field model by Cai \textit{et al.} \cite{Cai}. A significant effect of correlations was observed on nuclear symmetry energy $E_{sym} (\rho)$. Consequently, numerous experimental and theoretical efforts are currently underway and many others are in the planning stage.\\

An extension of traditional SRC and EMC measurements at Jefferson Lab is already envisaged in proposals E12-06-105 and E12-10-008 \cite{Arrington} in which a wide range of targets having different n/p ratios will be used with 12 GeV electron beam. These experiments aim to resolve the detailed relationship between the EMC effect and SRC phenomena along with the possible measurement of 3N SRCs at $Q^2$ values from 2 GeV${}^2$ to 20 GeV${}^2$, for Bjorken parameter $x>$1.  The measurements of inclusive electron scattering cross-sections on several nuclei, including first measurements on ${}^{10}$B and ${}^{11}$B targets with 10.6 GeV beam, have already been carried out and additional runs are planned in near future. These measurements are expected to quantify the role of local nuclear density in quark modification of nucleons inside nuclei resulting in the observed EMC effect. Another proposal, named as CaFe experiment, aims to measure SRCs in light and mid-mass nuclei \cite{Hen2} and to study their dependence on nuclear isospin asymmetry which has been approved by the Jefferson Lab. SRC probability measurements for targets ${}^2$H, ${}^4$He, ${}^9$Be, ${}^{10,11}$B, ${}^{12}$C, ${}^{28}$Si, ${}^{40}$Ar, ${}^{40,48}$Ca, ${}^{48}$Ti and ${}^{54}$Fe are expected to quantify the role of neutron-proton asymmetry in deciding the abundance of nn, pp and np SRCs in nuclei.\\

The earlier high impact SRC measurements at Jefferson Lab were reported with relatively small data samples consisting of a few hundred exclusive $(e,e^\prime pN)$ events. A re-run of these measurements with the CLAS12 spectrometer aiming for much better statistics at 4.4 GeV and 6 GeV beam energy with ${}^{2}$H, ${}^{4}$He, ${}^{12}$C, ${}^{28}$Si, ${}^{40,48}$Ca, ${}^{120}$Sn and ${}^{208}$Pb targets has been approved (PR12-18-003) \cite{Ashkenazy}. The experimental results from these measurements are expected in 2022. Another interesting experiment where real photons will be used as SRC probes has been proposed using the GlueX detector at Hall D of Jefferson Lab \cite{Alikhanian}. This would provide an alternative and independent method of measuring the dominance of np SRCs in nuclei.\\

 The exploration of the underlying relation between the EMC effect and high momentum nucleon pairs in SRCs will be one of the main physics objectives of the future Electron-Ion Collider (EIC) facility at BNL \cite{Hauenstein}. This quantification can be carried out by simultaneous measurement of scattered beam electrons after deep inelastic collision from nucleons along with their SRC paired counterparts ejected in the interaction event. The robust design of the proposed EIC detector \cite{Hauenstein} would allow for the tagged measurement of the outgoing neutrons and protons along with the reaction fragments produced in DIS events. \\
 
In addition to direct SRC measurements using energetic $e^-$ and p beams, many inverse kinematic experiments are being planned at JINR and GSI due to their unique ability for suppression of ISI/FSI using forward moving fragment tagging. A pioneering inverse kinematic SRC measurement made by BM@N collaboration using ${}^{12}$C beam at JINR \cite{Patsyuk} has been already discussed in Sec. 2. The upgrade of BM@N set up is ongoing and aims to improve the momentum resolution of outgoing fragments with higher acceptance angles. Exclusive SRC measurements with much higher statistics are expected in 2022-23 in which recoiling neutrons will also be measured along with other reaction products. This measurement is expected to shed light on the role of nuclear shells (s and p for ${}^{12}$C) on short-range NN correlations and possibly on 3N nuclear correlations.  \\

The planned SRC measurements using inverse kinematic measurements at FAIR (GSI) provide a unique opportunity for investigation of NN correlations in short-lived nuclei having extreme n/p ratio. A fully exclusive investigation of SRC inside neutron rich unstable nuclei such as ${}^{16}$C is being planned by Reactions with Relativistic Radioactive Beam (R${}^3$B) collaboration at GSI \cite{Duer2}. The scattering of ${}^{16}$C beam (up to energy of 1.25 GeV/u) with liquid hydrogen will be used to investigate SRCs. In past SRC measurements, the neutron detection has been a bottle-neck issue due to low neutron detection efficiency along with poor energy and position resolution. This has been overcome by R${}^3$B collaboration by constructing a segmented NeuLAND neutron detector \cite{Neuland} which would have about 90$\%$ neutron detection efficiency with good position, energy and timing resolution. The FAIR infrastructure allows SRC investigation in other extremely neutron rich nuclei like ${}^{24}$O, ${}^{52}$Ca, ${}^{70}$Ni and ${}^{215}$Pb which could be instrumental in understanding the NN correlations in neutron rich nuclei across the nuclear chart.  \\

As discussed in previous sections, NN correlations affect neutrino-nucleus interactions in a significant way and this effect is being included in recent neutrino-nucleus interaction models \cite{Khac}. The future neutrino experiments would shed more light on their effect on various neutrino-nucleus interaction channels. Further, NN correlations would be a deciding factor for outcome of numerous experiments like $(p,3p)$ knockout measurements at intermediate energy \cite{Frotscher} where  two sequential pp collisions appear to account most of the cross-section indicating a low probability of two-proton correlations in neutron rich nuclei. The extension of $(p,pd)$ measurement to mid- and heavy-mass nuclei may quantify the roll of tensor interaction in NN correlations as observed for ${}^{16}$O \cite{Tara} recently. Apart from further development of many existing models for NN correlations, as discussed in last two sections, some exciting ideas are being proposed by theorists. A recently proposed di-quark model by Rittenhouse West \cite{West} is gaining traction since it may provide a microscopic origin of NN correlations, dominance of $np$ SRCs and may shed light on EMC-SRC connection. \\

After establishing the presence of short-range correlations on firm footing, their implications on nuclear observables are being investigated by many groups \cite{Miller, Paschalis, Cosyn}. Miller \textit{et al.} investigated the influence of short-range correlations on nuclear properties through chiral dynamics \cite{Miller}. Recently, short-range correlations are identified as possible cause for the quenching of spectroscopic factors for direct reactions \cite{Paschalis}. The influence of these correlations on super-allowed $\bbeta$-decay is computed by Condren \textit{et al.} and a significant effect is reported \cite{Condren}. The formation of SRCs inside neutron rich nuclei would reduce their skin thickness and a $\sim$10$\%$ reduction in neutron skin is possible for ${}^{48}$Ca nucleus \cite{Cosyn}.

\section{Summary} 

In the present work, a review of the current status of a wide range of different experimental and theoretical works relating to two-nucleon correlated structures inside nuclei has been presented. Often, these investigations were carried out in separate self-contained investigtions in different time periods over the last 75 years. It is now becoming apparent that there are huge similarities and synergies between these studies of different reactions. In addition, it is clear that theoretical treatments of these effects have evolved along rather similar parallel lines, leading to a deeper understanding of the role and importance of NN correlations. While huge progress has been made over the past 70 years, there remain a large number of important questions to be answered by further investigation and study. Hopefully, the present review of these experimental and theoretical efforts will prove useful for a growing nuclear physics community aiming to understand the subtleties of 2N and 3N correlations across the nuclear chart. \\

$\bf{Acknowledgements}$

This work has been supported by grant ST/L00478X/1 from the UK Science and Technology Facilities Council.	 	


\end{document}